\documentclass[prb,aps,showpacs,citeautoscript, superscriptaddress,amsmath,amssymb,floatfix,twocolumn,dvipsnames,longbibliography]{revtex4-1}
\usepackage{graphicx}
\usepackage{xcolor}
\usepackage{ytableau}
\usepackage[colorlinks,bookmarks=false,citecolor=blue,linkcolor=red,urlcolor=black]{hyperref}
\usepackage{amsfonts}
\usepackage{amsmath}
\usepackage{times}
\usepackage{amssymb}
\usepackage{changes}
\usepackage{blkarray}
\usepackage{relsize}

\ytableausetup{smalltableaux} 

\setcitestyle{super}

\usepackage{color}
\definecolor{lightpink}{RGB}{255, 187, 218}
\definecolor{lightblue}{RGB}{130, 215, 255}
\definecolor{lightgrey}{RGB}{220, 220, 220}
\definecolor{lightred}{RGB}{255, 155, 155}
\definecolor{lightteal}{RGB}{143, 255, 243}
\definecolor{lightindigo}{RGB}{191, 181, 255}
\definecolor{lightgreen}{RGB}{185, 255, 179}
\definecolor{lightpurple}{RGB}{237, 179, 255}

\definecolor{c1}{RGB}{144,12,63}
\definecolor{c2}{RGB}{199,0,57}
\definecolor{c3}{RGB}{255,87,51}
\definecolor{c4}{RGB}{255,195,0}
\definecolor{c5}{RGB}{218,247,166}

\begin{document}

\title{Density Matrix Renormalization Group simulations of the SU(N) Fermi-Hubbard chain implementing the full SU(N) symmetry  via Semi-Standard Young Tableaux and Unitary Group Subduction Coefficients.}

\date{\today} 
\author{Pierre Nataf}
\affiliation{Laboratoire de Physique et Mod\'elisation des Milieux Condens\'es, Universit\'e Grenoble Alpes and CNRS, 25 avenue des Martyrs, 38042 Grenoble, France}

\begin{abstract}
We have developed an efficient method for performing density matrix renormalization group (DMRG) simulations of the SU(N) Fermi-Hubbard chain, fully leveraging the SU(N) symmetry of the problem. This method extends a previously developed approach for the SU(N) Heisenberg model and relies on the systematic use of the semi-standard Young tableaux (SSYT) basis in a DMRG algorithm "\`a la White".  Specifically, the method aligns the site-by-site growth process of the infinite-size part of the DMRG, in its original formulation, with the site-by-site construction of the SSYT (or Gelfand-like) basis, based on the chain of unitary subgroups $U(1)\subset U(2) \subset U(3) \subset U(4)\cdots $.
We give special emphasis to the calculation of the symmetry-resolved reduced matrix elements of the hopping terms between the left and the right block, which makes direct use of the basis of SSYT and of the Gelfand-Tsetlin coefficients, offering a computational advantage in scaling with $N$ compared to alternative methods that rely on summing over Clebsch-Gordan coefficients.
Focusing on the model with homogeneous hopping between nearest neighbors, we have calculated the ground state energy as a function of $U$, i.e the atom-atom interaction amplitude,  up to $N=6$ for filling $1/N$ (one particle per site in average), and for one  atom (resp. hole) away from filling $1/N$. It allows us to compute the charge gaps, and
to estimate in the thermodynamical limit, the critical value $U_c$ (separating the Mott insulator from the metallic  phase), which is proved to be finite (i.e. $U_c>0$) and to increase with $N$.
  Central charges $c$ are also extracted from the entanglement entropy using the Calabrese-Cardy formula, and are consistent with the theoretical predictions: $c=N-1$, expected from the SU$(N)_1$ Wess-Zumino-Witten CFTs in the spin sector for the Mott phase,  and $c=N$ in the metallic phase, reflecting the presence of one additional (charge) gapless critical mode.
  
\end{abstract}

\maketitle

 \section{Introduction}
Recent advancements in ultracold atoms have enabled experimentalists to model strongly correlated systems with increasing precision and sophistication \cite{Esslinger_2010,Mazurenko_2017,Schafer_2020}.
For example, degenerate gases of strontium and ytterbium loaded in optical lattices
have been used to engineer the SU(N) Fermi-Hubbard models (FHM) \cite{takahashi2012,Pagano2014,Scazza2014,Zhang2014,Hofrichter_2016,Becker_2021,taie2020observation,Fallani_2022,Pasqualetti_2024}. 
This is a generalization of the famous SU(2) FHM, which is crucial for the understanding of high-temperature superconductors and other quantum materials\cite{Hubbard_1963, Gutzwiller_1963,Scalapino2012Oct,Arovas2022Mar,Qin2022Mar}.
Such a generalization, first introduced as a theoretical tool to provide us with an approximation scheme valid in the large $N$ limit \cite{affleck_exact_1986,Affleck_1988,Rokhsar_1990,Marder_1990}, was later explored at finite $N$
in condensed matter, for, e.g., the study of some transition metal oxydes \cite{Li_1998,Yamada_2018}, graphene's SU(4) spin valley symmetry \cite{Fischer_2011}, Moir\'e bilayer graphene \cite{zhang2021} and in cold atoms (resp. molecules) on optical lattices \cite{wu_exact_2003,Honerkamp2004,Wu_review_2006,Cazalilla2009,gorshkov_two_2010,Capponi_annals_2016,Ibarra_Garcia_Padilla_2024,Mukherjee_2024},
where $N$ can be as large as $10$ (resp. $36$), and where the SU(N) symmetry is (quasi-)exact.
By the way, the variety of the color patterns that might emerge on a given lattice in the SU($N>2$) Mott phases is richer than the usual checkerboard pattern for $N=2$ on the square lattice at half-filling \cite{mambrini2003,bauer2012,nataf2014,Feng_2023}.
Incidentally, in the quest to identify conditions under which these systems could host more exotic phases than the $N-$colors Neel Antiferromagnetic states, such as SU(N) chiral spin liquids in two-dimensional system \cite{Schroeter_2007,hermele2009,Nielsen_2013,Kumar_2015,natafPRL2016,Chen_Hazzard_2016,Boos_2020,zhang2021,Chen_2021} or SU(N) Symmetry Protected Phase in one dimensional systems \cite{Gu2009,Chen_2010,Fidkowski_2011,nonne2011,nonne2013,quella2012,Quella2013_string,Quella2013_phases,Tanimoto_2015}, theoreticians have quickly encountered some challenges.

In fact, for most of the SU(N) models, there is no reliable analytical treatment. Important exceptions are provided by the SU(N) Heisenberg (resp. FHM) Hamiltonian with uniform interaction (resp. hopping) between nearest neighbors on a chain, for which Sutherland\cite{Sutherland1975}  (resp. Lieb and Wu\cite{Lieb_1968}) found the Bethe ansatz solution for any $N$(resp. $N=2$).

Otherwise, theoreticians should rely on numerical methods, and except in the cases  where there is no sign problem so that they can use Quantum Monte Carlo\cite{capponiQMC,Assad_2013,Demidio_2015,Demidio_2016,Wang_2014,Xu_2018}, they must often struggle against the explosion of the size of the Hilbert space, so that they are either limited to very small system sizes \cite{Nataf_2025}, either limited to small $N$ ($N=5$  being often already considered as large).
In fact, for the SU(N) FHM, for each added fermion, the dimension of the full Hilbert space is multiplied by $2^N$, which already gives $64$ for the (experimentally relevant) SU(6) fermions.

One strategy developed to overcome this difficulty is to implement the full SU(N) symmetry, which consists in working not in the full Hilbert space, but in a SU(N) symmetric sector invariant under the Hamiltonian, and corresponding to an SU(N) irreducible representation (irrep), usually labelled by a Young Diagram (YD)\cite{rutherford,cornwell,chen}.
It leads to a dramatic reduction of the number of linearly independent many-body wave-functions over which one should look for the states of smallest energies of the Hamiltonian.
For instance, for the exact diagonalization (ED) of the Heisenberg SU(6) model on a $L=12$-sites lattice, for the fundamental irrep at each site (of dimension $N=6$ since there are $N=6$ flavors),
the dimension of the full Hilbert space is $N^L\equiv 6^{12}\approx 2 \times 10^9$, while the dimension of the SU(6) singlet subspace (i.e the sector made of wave-functions invariant under local SU(6) transformations), is only $132$ \cite{nataf2014}, several order of magnitudes less.
For the SU(6) FHM at filling $f=1/6$, on the same $L=12$-sites cluster, going into the SU(6) singlets subspace is even more profitable, with a reduction of $13$ orders of magnitudes\cite{Botzung_2023_PRL}!

Interpreting the Hamiltonian of the SU(N) FHM (resp. Heisenberg model) as an element of the algebra of the unitary (resp.  permutation)  group,
we were able to implement the SU(N) symmetry in an ED algorithm using as a convenient basis, the set of semi-standard Young tableaux\cite{Botzung_2023_PRL,Botzung_2023} (resp. standard Young tableaux \cite{nataf2014}).
In this framework, the algebra of a group is seen as a part of a chain of subalgebras with natural embedding \cite{molev2002}. For $S_L$, the group of permutations of $L$ elements ($L$ is the number of sites of the cluster under consideration), the embedding:
\begin{align}
S_1 \subset S_2 \subset S_3 \subset \cdots \subset S_{L-1} \subset S_{L},  \label{embedding_permutation}
\end{align}
is at the heart of the construction (by A. Young \cite{Young6,rutherford}) of the basis of standard Young tableaux (SYT) for the irrep of the permutation group, and of the calculation of the matrix elements of the elementary generators of the permutation group (i.e permutation between consecutive $k$ for $1 \leq k \leq L$).

 \begin{figure*} 
\centerline{\includegraphics[width=1\linewidth]{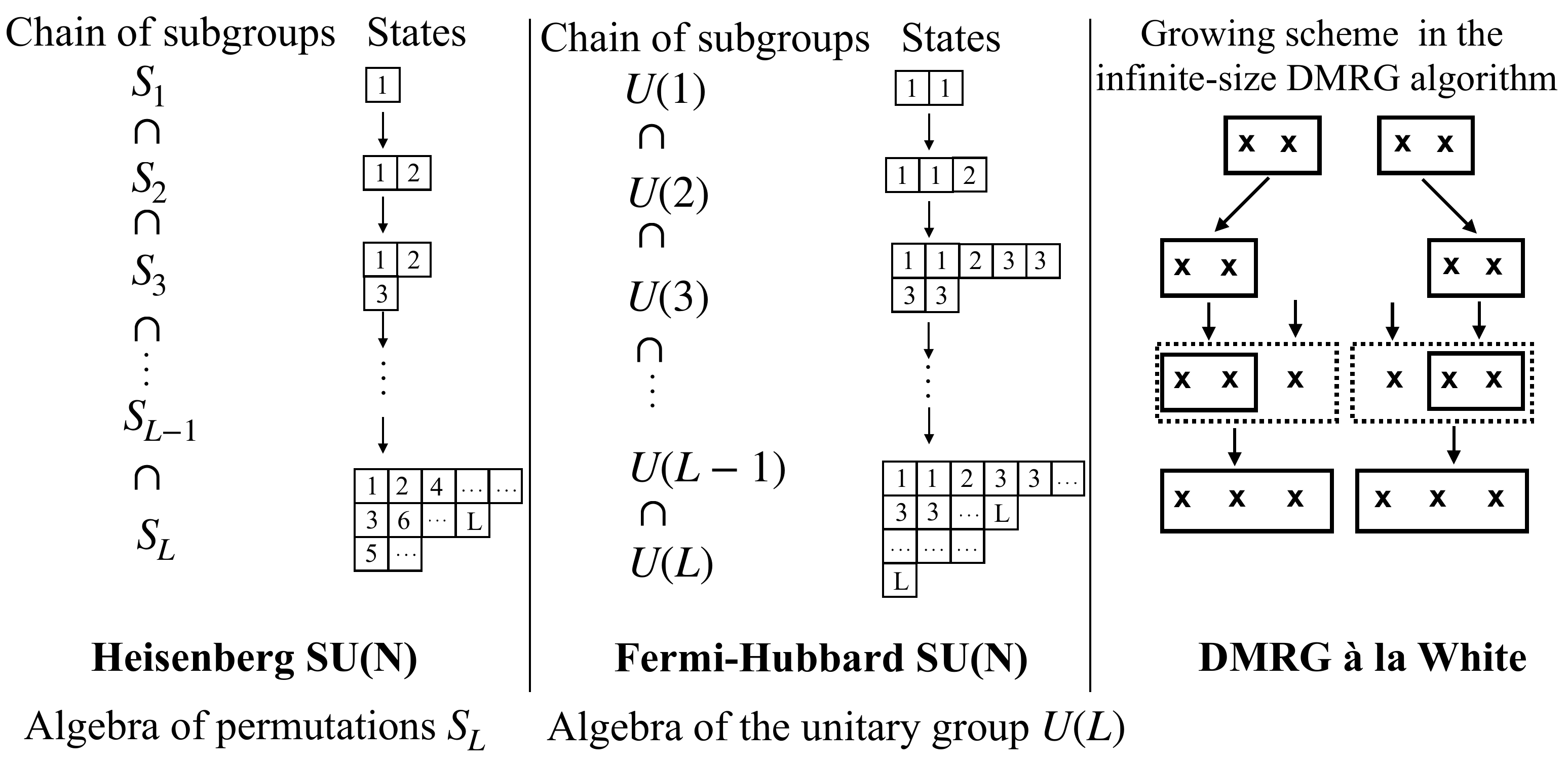}}
\caption{\label{subrgoup_chain_dmrg} Analogy between the site after site construction of the SU(N) symmetry-resolved states relevant for the Heisenberg SU(N) model (left)\cite{nataf_density_2018},
the Fermi-Hubbard model (middle)\cite{Botzung_2023_PRL}, and with the site after site growing process in the original version of the Density Matrix Renormalization Group (DMRG\cite{DMRG_white}) \`a la White (right).}
\end{figure*}

There is an interesting structural analogy between this site-after-site construction and the site-after-site growing process of the blocks in the Density Matrix Renormalization Group (DMRG) algorithm \cite{DMRG_white,whiteprb1993}
in the S. White's original formulation (cf. Fig \ref{subrgoup_chain_dmrg}), that we took advantage of in \cite{nataf_density_2018}, to implement the full SU(N) symmetry in a DMRG algorithm.
Such an implementation allowed us to bypass the calculation (and the painful storage) of the SU(N) Clebsch-Gordan coefficients (CGCs)\cite{alex2011}, and to address relatively large $L$ (up to several hundreds),
and large $N$ (up to $N=8$), with good precision ($6$ to $12$ digits, depending on N) for the ground state energies of the Heisenberg SU(N) model on a chain with fundamental irrep on each site, as compared to the Bethe ansatz solutions, with $m=1000$ states kept\cite{nataf_density_2018}.
 It also allowed us to solve some open physical problems around the numerical demonstration of the generalization of the Haldane's conjecture\cite{haldane_nonlinear_1983,haldane_continuum_1983} to SU(3)
 \cite{Affleck_1988,
rachel2007,
rachel2009,
bykov_geometry_2013,
Lajko_2017,
Tanizaki_2018,
Yao_2019,
wamer_generalization_2020}.
In particular,  we numerically proved\cite{Nataf_2021} that the Heisenberg model with a two-boxes symmetric irrep at each site (of dimension $6$) belongs to the universality class of the SU$(3)_1$ Wess-Zumino-Witten (WZW) conformal field theory (CFT)~\cite{affleck1986,ziman1987,Affleck_1988,Tanizaki_2018,Ohmori_2019} in agreement with the field theory predictions \cite{Lecheminant2015b}, and we demonstrated using intensive DMRG simulations \cite{gozel_2020} that the model with a three-boxes symmetric irrep at each site (of dimension $10$) was gapped\cite{Fromholz_2020}. In this latter case, by extrapolating our finite-size results, we were also able to obtain a $6$-digits value for the ground state energy in the thermodynamical limit, which was later confirmed (on top of the presence of the gap) \cite{Devos_2022} with variational uniform matrix product states (VUMPS) \cite{VUMPS_2018}.
 
 It is the purpose of the present paper to generalize those ideas to the SU(N) FHM, to have a DMRG algorithm to address the FHM on chains with the full SU(N) symmetry.
 Our protocol will be based on the embedding for the unitary groups \cite{molev2002} (cf. Fig \ref{subrgoup_chain_dmrg}):
\begin{align}
U(1) \subset U(2) \subset U(3) \subset \cdots \subset U(L-1) \subset U(L), \label{embedding_unitary}
\end{align}
which allowed Gelfand and Tstelin to build the representation of the unitary group in the basis of what is now called Gelfand-Tstelin patterns\cite{Gelfand_1950}, which are equivalent to the semi-standard Young tableaux (SSYT), that we have used in \cite{Botzung_2023_PRL} for ED.
In particular, we will show that our approach, which relies on the use of the {\it subduction coefficients} for the unitary groups \cite{chen} is significantly more efficient for large N than conventional CGCs-based methods.

The paper is organized as follows.

%

In section \ref{method}, we describe the structure of the DMRG code which is based on the original formulation of DMRG by S. White\cite{DMRG_white}, and which relies
on symmetry-resolved (multiplets) states written in the basis of SSYT for the implementation of the SU(N) symmetry. 
In particular, a part (cf subsection \ref{reduced_matrix_element}) is devoted to the calculation of the hopping term between the left and the right block, where the subduction coefficients play a key role.
In section \ref{study}, we apply our formalism to the one-dimensional SU(N) FHM with one particle per site on average (filling $1/N$), with uniform hopping between nearest neighbors. If well addressed through bosonization \cite{assaraf1999}, such a model remains the subject of a numerical controversy, about the presence of a finite metallic phase for positive $U$ and $N>2$ \cite{assaraf1999, manmana2011,Szirmai_2007}.
By addressing the chain with one particle (resp. hole), away from filling $1/N$ and for system sizes $L$ or number of colors $N$ larger than previously studied, we were able to calculate precisely the critical value $U_c$ separating the metallic from the insulating phase for $N=3,4$ and $N=6$. 
At the end of section \ref{study}, we also show the entanglement entropy and the central charges
 and demonstrate some good agreement with the expected SU$(N)_1$ CFT behavior for the spin sector,  with the presence of an additional critical mode in the metallic phase for $0\leq U < U_c$ \cite{assaraf1999}.
 
Finally, conclusions are presented, and detailed appendices containing several developments are provided to give interested readers the opportunity to reproduce the technical methodology.

\section{DMRG with implementation of the full SU(N) symmetry}
\label{method}
\subsection{Model and Description of the DMRG code}
\begin{figure} 
\centerline{\includegraphics[width=\linewidth]{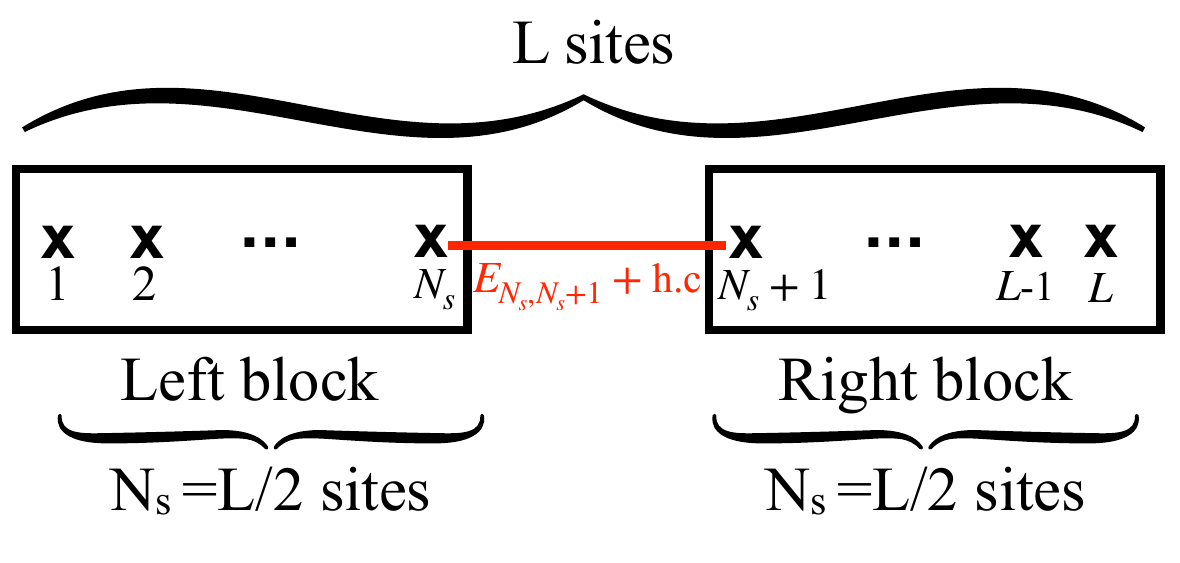}}
\caption{\label{sketch_systeme} System under consideration: open chain with $L=2 N_s$ sites. $N_s$ is the number of sites of both the left and the right blocks
which interact through the hermitian hopping $E_{N_s,N_s+1}+E_{N_s+1,N_s}$ where $N_s$ is the index of the site at the very right (resp. left) of the left (resp. right)
block.
}
\end{figure}

We describe here the DMRG algorithm that we use to study the SU(N) Fermi-Hubbard model (FHM)
whose Hamiltonian is:
\begin{equation}
\label{eq: Hamiltonian}
H = \sum_{\langle i,j \rangle}  \Big{(}  -t_{ij} E_{i,j}+ \text{h.c} \Big{)} +  \sum_{i=1}^L  \frac{U_i}{2} E_{i,i}^2, 
\end{equation}
where the SU(N) invariant hopping terms read
\begin{equation}
\label{eq: Hopping}
E_{i,j}=E_{j,i}^{\dag}=\sum_{\sigma=1}^N c^{\dag}_{i,\sigma}c_{j,\sigma}, 
\end{equation}
where the $\sigma$ are the color (or flavors) indexes.
The $t_{ij}$ in Eq. \eqref{eq: Hamiltonian}  are the hopping amplitudes between sites $i$ and $j$ and  $U_i$ is the local on-site density-density interaction for site $i=1 \cdots L$. 
For the numerical applications in the section \ref{study}, we will consider uniform hoppings between nearest neighbors: $t_{ij}\equiv t > 0$ if  $j=i+1$ (and $t_{ij}= 0$ if $\vert i-j \vert >1$), and with uniform positive interaction  $U_i \equiv U > 0 \,\,\forall i=1 \cdots L$, but non uniform parameters could be implemented within the very same code. We will focus on a L-sites one-dimensional chain with open boundary conditions (OBC) (cf Fig. \ref{sketch_systeme}), 
but the algorithm developed below could be adapted to other geometries or hoppings (rings, ladders, hoppings between next-nearest neighbors, etc...).

\begin{figure} 
\centerline{\includegraphics[width=\linewidth]{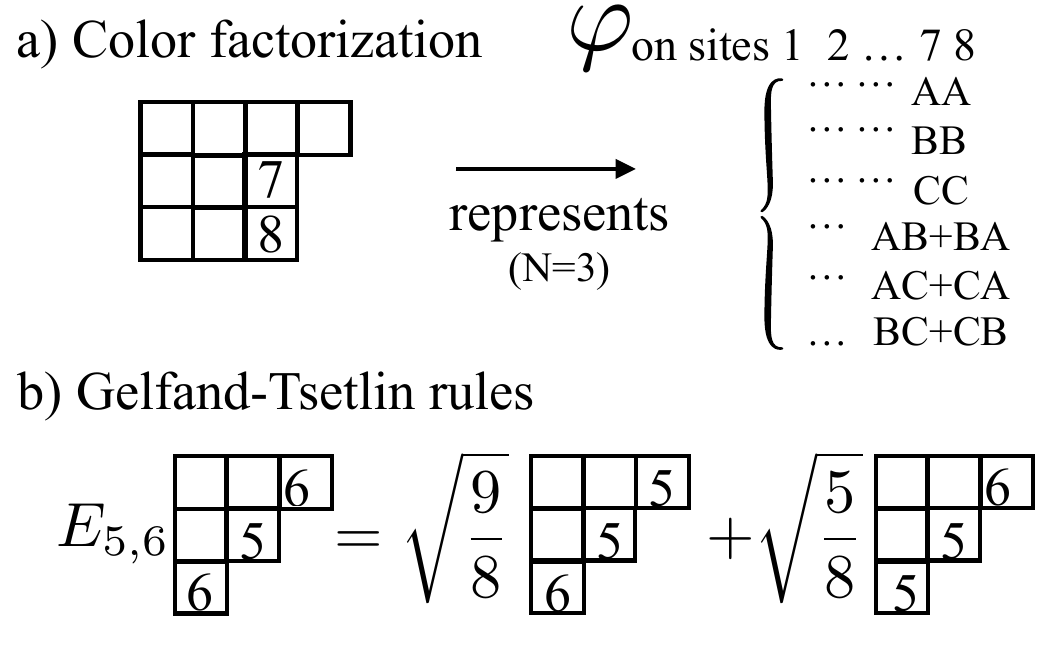}}
\caption{\label{sketch_summary_rules} a) Illustration of the color factorization at work in the use of the SSYT basis. Here, with $N=3$ colors named $A$, $B$ and $C$, 
the same SSYT can represent $6$ different class of wave-functions $\varphi$, which all live in the irrep $\bar{\alpha}=[433]$, with one particle on site 7 and one particle on site 8, which are symmetric under the exchange (or nullifies the hopping) between sites 7 and 8.  The unfilled boxes are arbitrarily filled (following SSYT rules) with numbers $\leq 6$. 
b) To know the effect of the operators $E_{p,p+1}$ on a given SSYT, one just needs to know where the numbers $p$ and $p+1$ appear in such a SSYT, and to apply the GT rules (cf Appendix \ref{GT_rules_appendix} and \ref{selection_appendix}).
}
\end{figure}

To implement the full SU(N) symmetry to address the SU(N) FHM, we use the basis of SSYT and the Gelfand-Tsetlin (GT) rules for the matrix representation of the generators of the unitary group in each irrep (cf \cite{Botzung_2023_PRL}  and Appendix \ref{GT_rules_appendix} for a review), in a way that is similar to the protocole developed for the DMRG simulations of the SU(N) Heisenberg with SYT \cite{nataf_density_2018}.
The key idea behind the use of SSYT in our DMRG simulations is that  each SSYT represents multiple many-body states that differ in their colors content but are equivalent under the action
of the SU(N) invariant hopping terms $E_{i,j}$ (cf Eq. \ref{eq: Hopping}). 
 This principle, known as color factorization  \cite{Botzung_2023_PRL,Botzung_2023}, is illustrated in Fig. \ref{sketch_summary_rules} a.

To illustrate our method, in both the current section and in the Appendices where we send most of the technical details, we focus on the infinite size part of the DMRG in the White's formulation: the chain is divided into the left and the right block, (cf  Fig. \ref{sketch_systeme}).
Each block is incrementally grown from size $N_s=L/2$ to size $N_s+1$ (see Fig. \ref{sketch_systeme}), and to prevent the Hilbert space from becoming too large, we perform a density matrix truncation on both the irreps for each block, and on the states within each irrep.
 This ensures that the total number of states retained in each block does not exceed $m$, an input parameter (typically $m$ is several thousands in section \ref{study}), which corresponds to the {\it bond dimension} in matrix product states (MPS) formulation of the DMRG.
Then,  since each SSYT represents several states, the implementation of the SU(N) symmetry results in an effective larger number of states kept $m$. 

\begin{figure} 
\centerline{\includegraphics[width=1\linewidth]{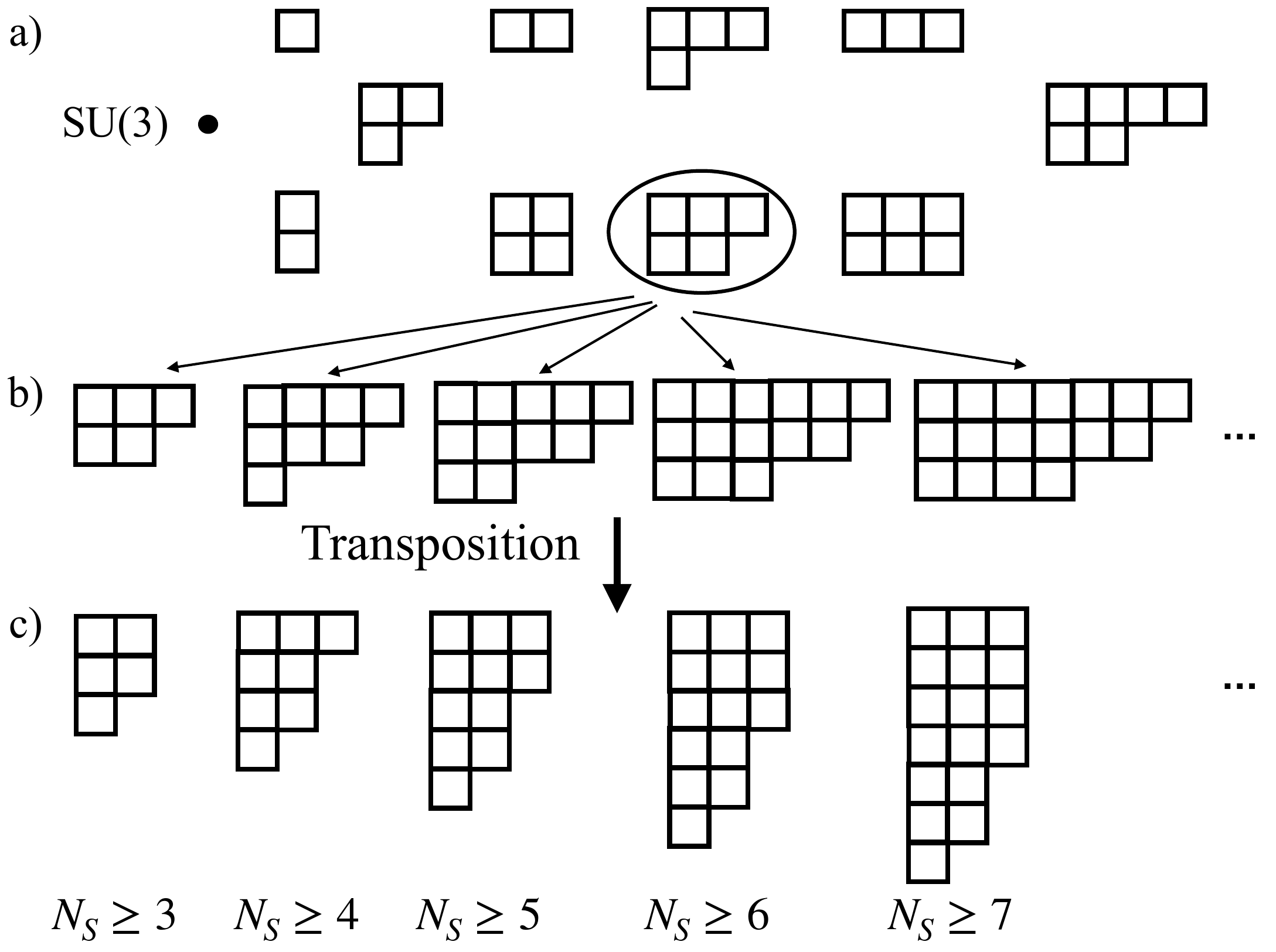}}
\caption{\label{sketch_irreps_truncation} a) Example of the truncation: we select the first $M=11$ SU(3) irreps according to the quadratic Casimir (cf Eq. \eqref{casimir}).
b) Equivalence of SU(N) irreps (here [3 2] for SU(3)) modulo added/withdrawn $N$-boxes columns. One can always choose the representation of an SU(N) irrep
through a Young Diagram with as much box as SU(N) fermions.
c) When transposed (rows changed into columns, and columns into rows), the YD represents an $U(N_s)$ irrep, where $N_s$ is the number of sites of each block, with at most $N$ columns
and $N_s$ rows.
}
\end{figure}

From the SU(N) symmetry perspective, each many-body state is characterized by its behavior under local SU(N) transformations, which determines its belonging to a specific sub-sector of the Hilbert space labeled by an SU(N) irrep, which generalizes the total spin $S$ for $N=2$.
%
%
An SU(N) irrep is {\it a priori} represented as an $N-1$ rows Young Diagram (YD) $\alpha=[\alpha_1,\alpha_2,\cdots,\alpha_{N-1}]$ with $\alpha_i$ the length of the $i^{\text{th}}$ row 
of the shape $\alpha$, satisfying $0 \leq \alpha_i \leq \alpha_{i-1}$ (for $i=2 \cdots N-1$) (cf Fig. \ref{sketch_irreps_truncation} a for some examples of SU(3) YDs) \cite{rutherford,cornwell,chen}.
 They can be characterized by the quadratic Casimir $C_2$ which depends on the shape of the SU(N) irrep $\alpha$ as \cite{schellekens,Botzung_2023}:
 \begin{align}
 \label{casimir}
 C_2=\frac{1}{2}\Big{\{}n(N-\frac{n}{N})+\sum_{i=1}^{N-1}\alpha_i^2-\sum_{j=1}^{j=\alpha_1}\bar{\alpha}_j^2\Big{\}},
 \end{align}
where the $\alpha_i$ ($i=1,..,N-1$) are the lengths of the rows and the $\bar{\alpha}_j$ ($j=1,..,\alpha_1$) are the lengths of the columns, and $n$ is the number of boxes of the irrep. When we add or withdraw a $N$-boxes column to a YD (from the left), it still represents the same SU(N) irrep, but it is useful when dealing with diagrammatic representation of SU(N) states (or multiplets)
like the SYT or the SSYT to choose the convention where there are as many fermions as boxes in the YD (cf Fig. \ref{sketch_irreps_truncation} b).

In our algorithm, like in \cite{nataf_density_2018}, we make a truncation over the SU(N) irreps: only the states living in the $M$ irreps of lowest quadratic Casimir $C_2$ (cf Eq. \ref{casimir}) will be considered (cf Fig. \ref{sketch_irreps_truncation} a),
where $M$ is an input parameter of the simulations  (typically $M=300$ to $420$ in section \ref{study}).
This is physically relevant for the antiferromagnetic phases, but may present challenges in the case of a ferromagnet or in the metallic phases: there, the convergence with the physical parameters shall be carefully controlled.
A benefit from this truncation is that the number of required group theory coefficients is finite (cf section \ref{reduced_matrix_element}), and can be computed once and for all before running the simulations.

Moreover, there are constraints on the shapes for the possible irreps at stage $N_s$: 
for a block of size $N_s$ sites, one can not have SU(N) YD with more than $N_s$ columns since we have fermions in the system, and more than $f \times N L$ boxes which is the total number of particles for the full chain, where $f$ is  the filling (one input of the algorithm) and $L=2N_s$ the number of sites of the entire chain.
We call $M_{N_s} \leq M$ the number of shapes satisfying these constraints at stage $N_s$.

\begin{figure} 
\centerline{\includegraphics[width=.8\linewidth]{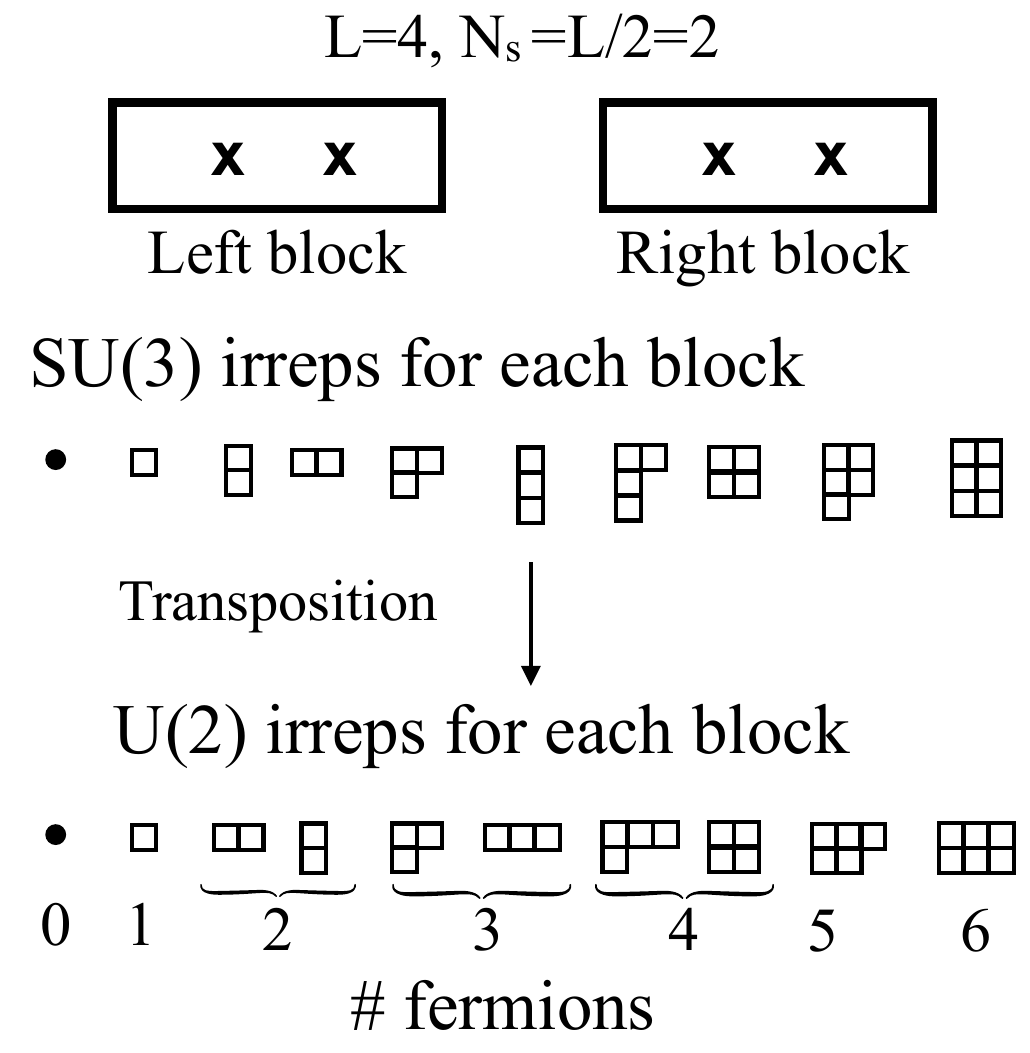}}
\caption{\label{sketch_irreps_constraints} Example of SU(N=3) irreps for each block of size $N_s=L/2=2$ ($L=4$ sites in the full chain).
The shapes of the SU(N) (resp. $U(N_s)$) irreps are constrained to have at most $N$ rows (resp. columns) and $N_s$ columns (resp. rows).
The number of box in each irrep is equal to the number of SU(N) fermions in each block.
}
\end{figure}

In addition, in order to use the basis of SSYT for a given set of SU(N) fermions on $N_s$ sites, one should {\it transpose}
the shapes $\alpha \rightarrow \bar{\alpha}$ (i.e transforming the rows into columns and columns into rows), to consider the shapes $\bar{\alpha}$ as $U(N_s)$ irreps, as shown in \cite{Botzung_2023_PRL,Botzung_2023}. Then, the constraints over the maximal number of rows/columns are transposed accordingly (cf Fig. \ref{sketch_irreps_truncation} and Fig. \ref{sketch_irreps_constraints}). Note that the transposition is an involution, i.e. $\bar{\bar{\alpha}} = \alpha$.

Finally, in the growing process: $N_s \rightarrow N_s+1$, the construction of the Hilbert space on each block depends on a state selection procedure based on entanglement entropy criteria requiring simple bookkeeping. In addition, as a byproduct of this selection, one can calculate the discarded weight $\mathcal{W}_d^{m,L}$, which represents the loss weight of the wave-function at size $L$ when we keep $m$ states.  
 During this process, the construction of the matrices representing the left or right block Hamiltonian is greatly facilitated by the use of SSYT and GT rules (cf Fig. \ref{sketch_summary_rules} b and Appendix \ref{GT_rules_appendix}), whose natural incremental structure (i.e. $N_s-1 \rightarrow N_s \rightarrow N_s+1\rightarrow$ etc ...) aligns seamlessly with the growth process in DMRG (cf also Fig. \ref{subrgoup_chain_dmrg}).
In the appendix section \ref{selection_appendix}, we provide a detailed description of the state selection, of the calculation of $\mathcal{W}_d^{m,L}$, and of the construction of the new matrices representing the SU(N) FHM in each block during the growing step $N_s \rightarrow N_s+1$.

\subsection{Hamiltonian on the superblock and reduced matrix element for the interaction between the left and the right block.}
\label{reduced_matrix_element}
\begin{figure} 
\centerline{\includegraphics[width=1\linewidth]{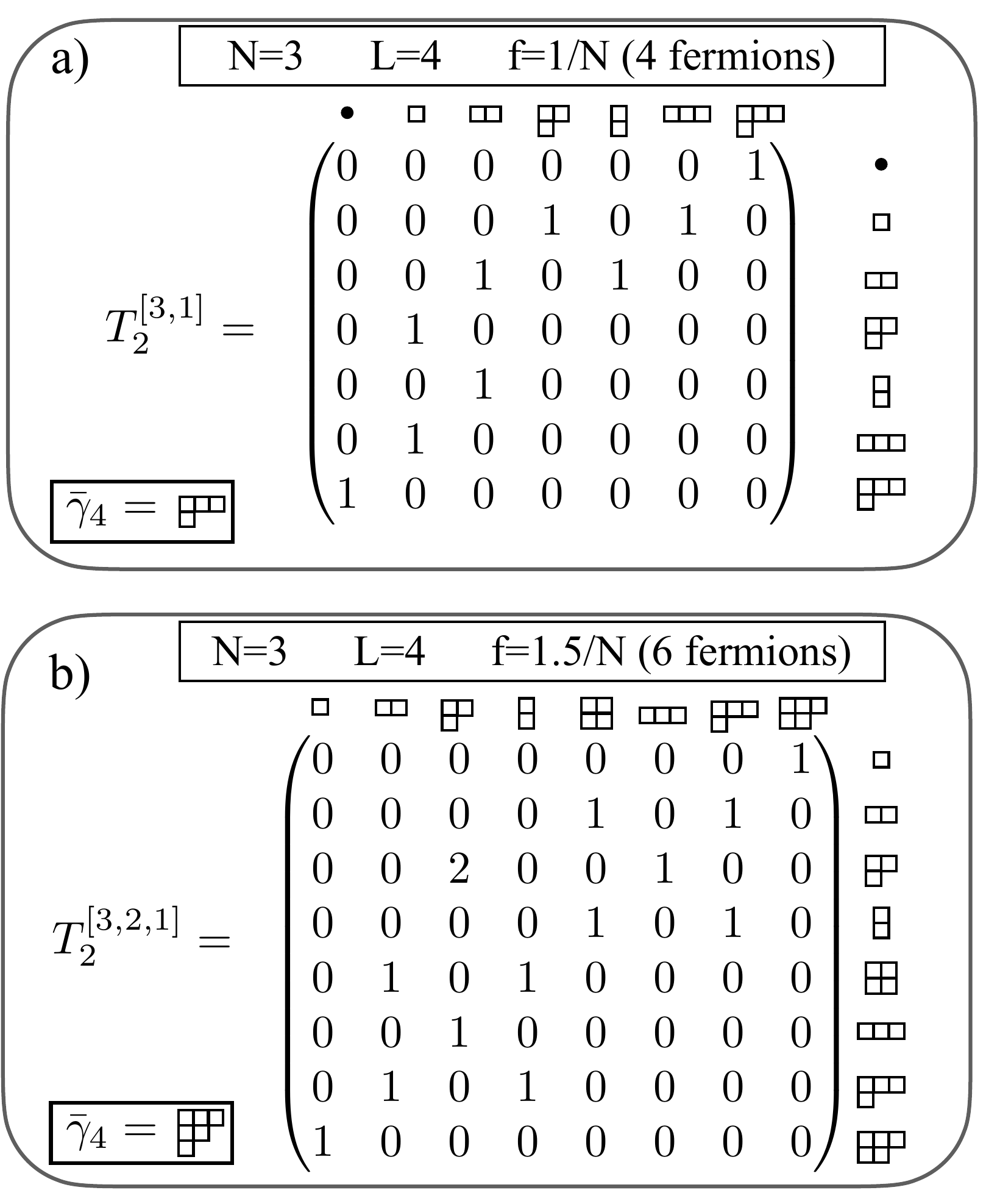}}
\caption{\label{matrix_bool} Examples of shapes $\bar{\gamma}_L$ representing some targeted irreps  and matrices $T^{\bar{\gamma}_L}_{N_s+1}$ for $N_s+1=2$  (i.e. $L=4$) for SU(3) and for the targeted irrep $\bar{\gamma}_L=[3 1]$ in a), and for the targeted irrep $\bar{\gamma}_L=[3 2 1]$ in b).
The shapes $\bar{\beta}_q$ (resp. $\bar{\beta}_{q'}$) for $q$ (resp $q'$)$=1,\cdots,M_{N_S+1}$ which label the columns (resp. lines) of the matrices  $T^{\bar{\gamma}_L}_{2}$ are the transposed Young Diagram of the shapes representing the SU(N) irreps for the left (resp. for the right) block made of $N_s+1=L/2(=2)$ sites (cf Fig. \ref{sketch_irreps_constraints}).
Then, the entries $T^{\bar{\gamma}_L}_{N_s+1}(q,q')$ are the multiplicities of $\bar{\gamma}_L \in \bar{\beta}_q \otimes \bar{\beta}_{q'}$ 
(see text for details)}
\end{figure}
After the state selection and the construction of the SU(N) FHM matrices on each block, one should calculate the matrix representing the Hamiltonian $H_L$ for the full chain at stage $N_s+1$.
 $H_L$ should be decomposed as:
\begin{align}
H_L=H_{N_s+1}^{\text{Left}}+H_{N_s+1}^{\text{Right}}-t(E_{N_s+1,N_s+2}+h.c),
\end{align}
where $E_{N_s+1,N_s+2}$ is the hopping term from the very first site of the right block to the very last site of the left block (cf Fig. \ref{sketch_systeme}), and where $H_{N_s+1}^{\text{Left}}$
(resp. $H_{N_s+1}^{\text{Right}}$) is the Hamiltonian of the SU(N) FHM for the left (resp. right) block of size $N_s+1$.
It is necessary to identify which target irrep $\bar{\gamma}_L$ one aims to obtain the ground state in.
If the goal is to target the absolute ground state, the target irrep depends on the specific parameters of the problem. In such cases, one can perform exact diagonalization (ED) on small chains to gain a general idea.

For instance, for $U \gtrsim \vert t \vert$, and for filling $1/N$ (one particle per site on average), the ground state of the SU(N) FHM is 
antiferromagnetic, living in the SU(N) singlet irrep $\gamma_L=[L/N,L/N, \cdots, L/N]$ (i.e $\bar{\gamma}_L=[N,N, \cdots, N]$, with $L/N$ rows) for $L$ multiple of $N$, and in the most antisymmetric one otherwise.

To build the superblock, one first calculate the $M_{N_s+1}\times M_{N_s+1}$  matrix $T^{\bar{\gamma}_L}_{N_s+1}$ whose coefficients $T^{\bar{\gamma}_L}_{N_s+1}(q,q')$ are  
 equal to the outer multiplicity of $\bar{\gamma}_L$ in  $\bar{\beta}_q \otimes \bar{\beta}_{q'}$, for $1\leq  q,q'\leq   M_{N_s+1}$, and
where the $\bar{\beta}_q$  and $\bar{\beta}_{q'}$ are among the $M_{N_s+1}$ shapes $\bar{\beta}$ for the left and the right block.
The rules to make the tensor product of two $U(N_s)$ irreps are reviewed in section \ref{tensor_product_appendix}.
Note that when $ \gamma_L \notin \bar{\beta}_q \otimes \bar{\beta}_{q'}$, $T^{\bar{\gamma}_L}_{N_s+1}(q,q')=0$. We give in Fig. \ref{matrix_bool} some instances of the matrix $T^{\bar{\gamma}_L}_{N_s+1}$ for $N=3$ for two different targeted irreps $\gamma_{L=4}$.

From $T^{\bar{\gamma}_L}_{N_s+1}$ and from the structure of the truncated Hilbert space for both the left and the right block (i.e. which state lives in which irrep), one can create the Hilbert space for the superblock, as well as the matrices representing $H_{N_s+1}^{\text{Left}}$ and $H_{N_s+1}^{\text{Right}}$ on it, as detailed in Appendix \ref{superblock_appendix}.

The last required step is then the calculation of the matrix $E_{\bar{\gamma}_L} \equiv E^{\bar{\gamma_L}}_{N_s+1,N_s+2}$, which represents the hopping term from the very first site of the right block to the very last site of the left block, i.e. $E_{N_s+1,N_s+2}$ on the sector $\bar{\gamma}_L$ (cf Fig. \ref{sketch_systeme}).

Let's consider four states $\vert \eta^{\bar{\beta}_{q_i}}_{i} \rangle$, each of them belonging to a sector labelled by the irrep $\bar{\beta}_{q_i}$ (for $i=1, \cdots 4$), and where both $\vert \eta^{\bar{\beta}_{q_1}}_{1} \rangle$ and $\vert \eta^{\bar{\beta}_{q_3}}_{3} \rangle$ (resp. $\vert \eta^{\bar{\beta}_{q_2}}_{2} \rangle$ and $\vert \eta^{\bar{\beta}_{q_4}}_{4} \rangle$) live in the Hilbert space of the left (resp. right) block.
First of all,  the coefficients $\langle \eta^{\bar{\beta}_{q_3}}_{3}  \vert \otimes\langle \eta^{\bar{\beta}_{q_4}}_{4} \vert E_{\bar{\gamma}_L} \vert  \eta^{\bar{\beta}_{q_1}}_{1} \rangle \otimes \vert \eta^{\bar{\beta}_{q_2}}_{2}\rangle$
will be zero unless that
  both $\bar{\gamma}_L\in \bar{\beta}_{q_1} \otimes \bar{\beta}_{q_2}$ and $\bar{\gamma}_L\in \bar{\beta}_{q_3} \otimes \bar{\beta}_{q_4}$, i.e $T^{\bar{\gamma}_L}_{N_s+1}(q_1,q_2),T^{\bar{\gamma}_L}_{N_s+1}(q_3,q_4)>0$.
 Secondly, on the basis of SSYT, the value of such a coefficient is determined by the positions of the indices of the very last added site (i.e $N_s+1$ in the shapes $\bar{\beta}_{q_1}$ and $\bar{\beta}_{q_3}$, and $N_s+2$ in the shapes $\bar{\beta}_{q_2}$ and $\bar{\beta}_{q_4}$). Thus, and analogously to the Heisenberg case, one can introduce the {\it shape and cross} notation for these coefficients, which
 keep track of the chain of sectors when making each block grow from $N_s$ sites to $N_s+1$ sites.
 In particular, we will work on the following example which is useful for $N\geq4$:
 \begin{align}
&\text{\raisebox{-2.4ex}{$\mathlarger{\mathlarger{\mathlarger{\mathlarger{\mathlarger{\mathlarger{\langle}}}}}}$}}  \ytableaushort{\,\,\,\times,\,\,\,,\,\times}\otimes   \ytableaushort{\,\,\,\,,\,\,\times,\,,\,}\text{\raisebox{-2.4ex}{$\mathlarger{\mathlarger{\mathlarger{\mathlarger{\mathlarger{\mathlarger{\vert}}}}}}$}} E_{\text{\small $\ytableaushort{\,\,\,\,,\,\,\,\,,\,\,\,\,,\,\,\,,\,\,\,}$}} \text{\raisebox{-2.4ex}{$\mathlarger{\mathlarger{\mathlarger{\mathlarger{\mathlarger{\mathlarger{\vert}}}}}}$}}  \ytableaushort{\,\,\,\times,\,\,\,,\,}\otimes   \ytableaushort{\,\,\,\,,\,\,\times,\,\times,\,}\text{\raisebox{-2.4ex}{$\mathlarger{\mathlarger{\mathlarger{\mathlarger{\mathlarger{\mathlarger{\rangle}}}}}}$}} =\nonumber  \\ &\langle \bar{\beta}_{q_3},l_3  \vert \otimes\langle \bar{\beta}_{q_4},l_4 \vert  E^{\bar{\gamma}_L}_{N_s+1,N_s+2} \vert \bar{\beta}_{q_1},l_1\rangle \otimes \vert \bar{\beta}_{q_2},l_2\rangle=\frac{\sqrt{3}}{2\sqrt{2}}  \label{coeff_ex}
\end{align}
 In the above example, $\bar{\gamma}_L=[4,4,4,3,3]$, $ \bar{\beta}_{q_1}=[4,3,1], \bar{\beta}_{q_2}=[4,3,2,1], \bar{\beta}_{q_3}=[4,3,2]$ and $\bar{\beta}_{q_4}=[4,3,1,1]$.
 The vectors $l_j$  are the indices of the rows where the cross appear in each $\bar{\beta}_{q_j}$ (for $j=1, \cdots 4$): $l_1=[1], l_2=[3,2], l_3=[3,1]$, and $l_4=[2]$. 
 In such a notation, the shapes without (resp. with) the boxes containing the cross represent the irrep when each block has $N_s$ (resp. $N_s+1$) sites.
 In particular, the coefficient in Eq. \ref{coeff_ex} involves nine different irreps (i.e. $\bar{\gamma}_L$ and the four times two irreps $\bar{\beta}_{q_j}$, with/without the cross for $j=1 \cdots 4$), so these coefficients are in close connection with the Wigner 9j coefficients \footnote{Such a connection goes beyond the scope of the present manuscript and might be studied in a future work}.
The calculation of the coefficient in Eq. \ref{coeff_ex} is made in details through the use of the basis of SSYT in Appendix \ref{subduction_coeff}. It necessitate in particular the {\it tensor product} of two SSYT on the targeted irrep $\bar{\gamma}_L$, which requires what is known as the subduction coefficients of the unitary groups \cite{chen}.
  Due to the truncation over the irreps, the number of such a coefficients is finite (cf Appendix \ref{subduction_coeff}).

The main result of the methodological part is that the SSYT-based method for the computation of the SU(N) symmetry-resolved reduced matrix elements of the interaction between the two blocks scales avantageously with $N$, as compared with alternative methods. These latter are often based on the direct use of the Wigner 9j or 6j or some related  (like the X symbols \cite{weichselbaum2020}) coefficients which naturally arise through the Wigner-Eckart theorem \cite{mcculloch2002,mcculloch2007}
The reason is that, apart from SU(2) where there are closed-form expression (cf Eq. (3.326) in \cite{Biedenharn_Louck_1981}), the calculation of SU(N) Wigner n-j (n=6 or 9) coefficients 
\cite{Kramer1968,Dang2024} relies most frequently on the summation over CGCs, whose time of computation is a polynomial in the dimensions of the two SU(N) irreps (before transposition) involved in the tensor products (cf Appendix \ref{tensor_product_appendix}).
In the example above, the SU(4) irrep $\bar{\bar{\beta}}_{q_1}=[3 2 2 1]$ (resp.  $\bar{\bar{\beta}}_{q_2}=[4 3 2 1]$) has dimension 15 (resp. 64), so that we can still survive with CGCs-based methods. However, 
 we are able to calculate:
\begin{align}
&\langle \bar{\beta}_{q_3},l_3  \vert \otimes\langle \bar{\beta}_{q_4},l_4 \vert  E^{\bar{\gamma}_L}_{N_s+1,N_s+2} \vert \bar{\beta}_{q_1},l_1\rangle \otimes \vert \bar{\beta}_{q_2},l_2\rangle= \sqrt{\frac{32}{27}} \label{coeff_ex2}, \\  
&\text{with} \hspace{2cm}  \bar{\gamma}_L=[6 6 6 6 6 6 6 2], \nonumber\\
 &\bar{\beta}_{q_1},l_1=[6 5 4 3 2 1],[5 1] \hspace{.5cm} \bar{\beta}_{q_2},l_2= [6 5 4 3 2 2 1],[6 4 1]\nonumber, \\
&\bar{\beta}_{q_3},l_3=[6 5 4 4 2 1],[5 4 1] \hspace{.5cm} \bar{\beta}_{q_4},l_4= [6 5 4 2 2 2 1],[6 1]\nonumber,
\end{align}
 in few seconds on a laptop. In particular, such a computation required the diagonalization of a matrix (similar to the matrix shown in Appendix \ref{subduction_coeff} Eq. \eqref{matrix_operator}, for the simpler example of Eq. \eqref{coeff_ex}) of dimension 164 to get the subduction coefficients. This is much smaller than the dimensions of the SU(6) irreps $\bar{\bar{\beta}}_{q_1}=[6 5 4 3 2 1]$ and $\bar{\bar{\beta}}_{q_2}=[7 6 4 3 2 1]$, which are respectively 32768 and 145530, which are way too large to allow us for the computation of the above reduced matrix element with summation over CGCs.

Finally, from $\mathcal{H}^{\bar{\gamma}_L}_{L}$, i.e the matrix representing $H_L$ on ${\bar{\gamma}}_L$, one calculates both the minimal energy and the ground state $ G^L_{\bar{\gamma}_L}$, which is the eigenvector of minimal energy. From $ G^L_{\bar{\gamma}_L}$, as shown in Appendix \ref{superblock_appendix},  we get the reduced density matrices in each irrep within each block, which are useful both to get the entanglement entropy $S(L)$ and to prepare the next step (i.e. $N_s+1 \rightarrow N_s+2$, cf Appendix \ref{selection_appendix}).

 \section{DMRG results for the SU(N) Fermi-Hubbard chain with open boundary conditions}
 \label{study}
As an application of our DMRG algorithm, we have numerically investigated the Metal-Insulator transition in the one-dimensional SU(N) FHM with OBC for $N=3,4$ and $N=6$ at filling $1/N$ for positive on-site interaction $U$ at T=0K.

From an experimental point of view, this question is important as the metal-insulator transition has not yet been observed to the best of our knowledge for $N>2$ on the 1D lattice (contrary to the cubic lattice \cite{Hofrichter_2016}) although the implementation of such a model for these physical parameters becomes realistic with the recent developments: SU(6) Ytterbium atoms were for instance loaded in a 1D chain at very low temperature  \cite{taie2020observation} (i.e. lowest temperature achieved $T/t \sim 0.1$). 
From the theoretical point of view, some fundamental questions about this transition at such a filling remain open for $N>2$.
To this extent, the situation at filling $1/N$ is different from the filling $1/2$ where the conclusion is established: a Mott transition occurs at $U_c=0$ like for $N=2$\cite{Szirmai_2005,Szirmai_2007,Capponi_annals_2016,nonne2011}.
 
 In fact, at filling $1/N$, the analytical bosonization procedure cannot tell us whether such a transition occurs at finite $U_c>0$ or not \cite{assaraf1999,Szirmai_2005} 
  but it can give important informations about the nature of the transition, and about the effective low energy theories in each phase on both sides of the transition \cite{assaraf1999}.
An estimate of $U_c$ for  $N>2$ relies then on numerics, and the previous studies gave opposite conclusions.
For instance, in \cite{assaraf1999,manmana2011} they got results consistent with a finite $U_c>0$, as opposed to \cite{Szirmai_2007} where the authors find $U_c \approx 0$.

It is therefore legitimate to imagine that DMRG simulations with full SU(N) symmetry implemented could yield either more accurate results (i.e., with smaller discarded weight $\mathcal{W}_d^{m,L}$ at fixed $L$), or results on longer chains (i.e., $L \sim 100$), or with a larger number of colors (up to $N = 6$), all avenues that could be helpful in resolving the controversy.

Our methodology is twofold: firstly, we calculate the charge gap $\Delta_c$ to fit $U_c$ (cf Eq. \eqref{KT_exp_law} below) and secondly, we characterize  the effective critical low energy theories in both the metallic and the insulating phases by computing the central charges through the Calabrese-Cardy formula (cf Eq. \eqref{calabrese_cardy} below).

We set the hopping amplitude $t\equiv 1$ in the following.

For $N>2$, the transition is expected \cite{assaraf1999} to be of Kosterlitz-Thouless (KT) type at $U_c \geq 0$,
with a charge gap $\Delta_c$ behaving (in the thermodynamical limit) as:
\begin{align}
\label{KT_exp_law}
\Delta_c=C_{KT}\text{exp}\Big{(}-\frac{G_{KT}}{\sqrt{U-U_c}}\Big{)},
\end{align}
for $U>U_c$ and zero otherwise.

\begin{figure*} 
\centerline{\includegraphics[width=1\linewidth]{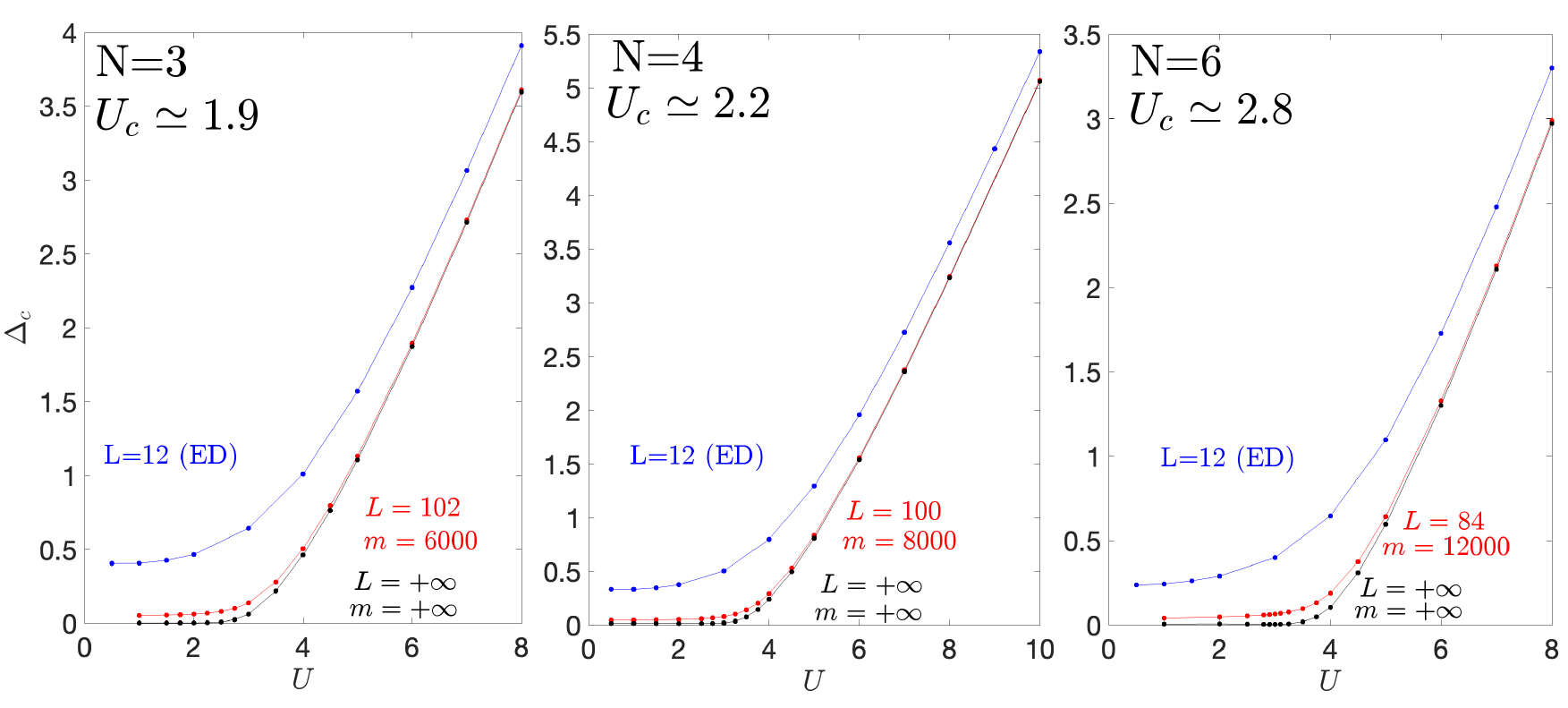}}
\caption{\label{gap_charges} Charge gaps $\Delta_c$ for $N=3$ (left), $N=4$ (middle), and $N=6$ (right), as a function of $U$. We calculated the ground state energies $E_0$ of the Hamiltonian in Eq. \ref{eq: Hamiltonian} for $t=1$ as a function of $U$, for the three doping $\delta=0, \pm1$, (i.e or fillings $f_L=1/N+\delta/(NL)$) corresponding to three different targeted irreps (cf Eq. \ref{charge_gap} for the definition of $\Delta_c$). For $L=12$ (in blue), we used ED, 
while for $L=L^{\Delta_c}_{\text{Max}}=102$ ($N=3$), $100$ ($N=4$) and $84$ ($N=6$), we used our DMRG algorithm with full SU(N) symmetry with $m=6000$  ($N=3$), $m=8000$ ($N=4$) and $m=12000$ states kept ($N=6$) (cf also Tab. \ref{table: table1} for some numerical values).
In black, extrapolated values (cf Fig. \ref{Figure_fitting} for the extrapolation procedure) in the limit $L\rightarrow +\infty, m \rightarrow + \infty$. See text for details. }
\end{figure*}

The (finite-size) charge gap is defined as: 
\begin{align}
\label{charge_gap}
\Delta_c=\sum_{\epsilon=\pm1}E_0(f_L=\frac{L+\epsilon}{NL})-2E_0(f_L=\frac{1}{N}),
\end{align}
 where $E_0(f_L)$ is the minimal energy for the $L$-sites chain with filling $f_L$.
In particular, $f_L=\frac{1}{N}$ means that there is one particle per site on average, while $f_L=\frac{L+1}{NL}$ (resp. $f_L=\frac{L-1}{NL}$) means one particle (resp. hole) away  from the filling $1/N$.
Introducing the doping $\delta=0, \pm1$, the total number of particles reduces to $f_LNL=L+\delta$. 
Note that the quantity $\Delta_c$, presented in Eq. \ref{charge_gap} as the definition of the charge gap in accordance with \cite{assaraf1999}, can instead be interpreted as the finite-size compressibility. Regardless of the terminology, this quantity serves in any case as an indicator of the metal-insulator transition.

From ED on small chains, we were able to scrutinize all the relevant irreps for the three different doping $\delta$ (or fillings $f_L$), and to infer that the ground state for $U\geq0.5$ should always live in the most antisymmetric $N$-rows and $f_LNL$-boxes YD, which will be the targeted shape $\gamma_L$ (before transposition).
We have also used the ED ground state energies $E_0$ and charge gap $\Delta_c$ for $L=12$ to benchmark our code, and we show these quantities for $N=2,3,4$ and $6$, for filling $1/N$ and  for $U=1$ and $U=5$ in Tab. \ref{table: table0}.
For $N=3$, $m=8000$ states kept and $M=300$ irreps, we have 10 (resp. 9) good digits for $E_0$ at $U=5$ (resp. $U=1$), while for $N=6$, $m=12000$ states kept and $M=420$ irreps, we have 7 (resp. 6) good digits at $U=5$ (resp. $U=1$).
It is a general trend that we have observed in all our simulations: as we will show below, for all values of $N$, $U=5$ (resp. $U=1$)  will be in the insulating (resp. metallic) phase, and the numerical convergence is better in the insulating phase than in the metallic phase, due to the Casimir-based truncation of the $M$ irreps (cf Fig. \ref{sketch_irreps_truncation} a).


 \begin{table}[h]
    \centering
\begin{tabular}{|c|c|c|c|c|} 
\hline
N&$E_0(U=1)$&$\Delta_c(U=1)$&$E_0(U=5)$&$\Delta_c(U=5)$\\
\hline
2 & -11.840637285901&0.53823&-5.535630158601&2.42139 \\
\hline
3&-15.376173634063&0.40623&-7.024399312653&1.57037\\
\hline
4& -16.722033360620&0.33088&-7.602319080277&1.29482\\
\hline 
6&-17.700163882249&0.24275&-8.029094742355&1.09706
\\
\hline 
\end{tabular}
\caption{Exact Ground state Energies $E_0$ (at filling $1/N$, i.e doping $\delta=0$) and charge gap $\Delta_c$ (cf Eq. \ref{charge_gap}) of the SU(N) FHM on the chain with OBC for L=12 sites calculated through Exact Diagonalization for $U=1$ and $U=5$. We have exactly diagonalized the Hamiltonian shown in Eq. \ref{eq: Hamiltonian} (from which we have withdrawn the constant $-LU/2$ for convenience) taking into account the full SU(N) symmetry using the method introduced in \cite{Botzung_2023_PRL} \label{table: table0} and reviewed in Appendix \ref{GT_rules_appendix}.}
\end{table}

For each value of $N$ and $U$ that we have considered,  we have performed the infinite size DMRG up to $L=L^{\Delta_c}_{\text{Max}}$ for 
the three different filling $f_LNL=L+\delta$ ($\delta=0, \pm1$), and for several (at least two) values of $m$, i.e. the number of states kept, and (unless otherwise specified)
for $M=300$ irreps kept.
The chosen values of $L^{\Delta_c}_{\text{Max}}$ represent a trade-off between computational time and having a sufficiently long chain to ensure a reliable extrapolation in the $L=+\infty$ limit (cf after).
For instance, for $N=4$, $L^{\Delta_c}_{\text{Max}}=100$, and we made simulations with $m=6000, 8000$ and $10000$ states kept.
We show in Tab. \ref{table: table1}, some DMRG results for the total ground state Energy $E_0(m,L)$ and discarded weight $\mathcal{W}_d^{m,L}$, at doping $\delta=0$, for various values of $N$ and $L$ as well as
 the charge gaps $\Delta_c(m,L)$ obtained from the three simulations (i.e $f_L=(1/N)+\delta/(NL)$ with $\delta=0, \pm1$, cf Eq. \ref{charge_gap}), for $U=1$ and $U=5$.
 This latter quantity is also plotted in red as a function of $U$ in Fig. \ref{gap_charges}.
Importantly, we made two different kinds of extrapolation. 
A first extrapolation gives our best estimate of the minimal energies  at fixed $L$. It is made by using the loss weight $\mathcal{W}_d^{m,L}$ (in the abscissa) obtained for various values of $m$: in the limit $m\rightarrow +\infty$, $\mathcal{W}_d^{m,L}\rightarrow 0$, and $E_0(m,L)\rightarrow E_0(m=+\infty,L)$, where $E_0(m=+\infty,L)$ is obtained through a linear fitting.
$E_0(m=+\infty,L)$  for $L=36,48$ and $84$ and for $U=1$ and $5$ and the largest values of $m$ used for the extrapolation (i.e.  $m_{\text{Max}}$)  are shown in Tab. \ref{table: table1}.

A second extrapolation is made at fixed $m$ in the thermodynamical limit $L \rightarrow \infty$, fitting the points $\Delta_c(m,L)$ with a function of the form $a+b/L+d/L^2$ to obtain
$\Delta_c(m,L=+\infty)=a$, as illustrated in Fig. \ref{Figure_fitting} in Appendix \ref{fitting_example}, for $N=6$. 

Then, we take the limit $m \rightarrow +\infty$ in $\Delta_c(m,L=+\infty)$, using the first kind of extrapolation (i.e with the discarded weight at $L=L^{\Delta_c}_{\text{Max}}$, as shown in the insets of Fig. \ref{Figure_fitting}), to get $\Delta_c(m=+\infty,L=+\infty)$ \footnote{Note that the two limits do not commute exactly, but the difference for  $\Delta_c(m=+\infty,L=+\infty)$  is irrelevant ($\lesssim 1e-3$) and does not change our estimate of $U_c$}. 
To make this extrapolation for many values of $U$, we considered simulations with $m_1$ and $m_2$ states kept, where $m_1$ and $m_2$ are tabulated in Tab. \ref{table: table1}.
 We show the resulting $\Delta_c(m=+\infty,L=+\infty)$ in black as a function of $U$ in Fig. \ref{gap_charges}. 
We fit $\Delta_c(m=+\infty,L=+\infty)$ with the function shown in Eq. \ref{KT_exp_law} to obtain our estimates of $U_c$ which read (cf also Fig. \ref{gap_charges}):
\begin{align}
\label{finite_U_c}
U_c(N=3) \simeq 1.9, \nonumber \\ 
U_c(N=4) \simeq 2.2,\\
U_c(N=6) \simeq 2.8.  \nonumber 
\end{align}
In spite of the good accuracy of our DMRG energies (error of the order of the discarded weight $\sim 1e-5$ at worst, depending on $N$, $L^{\Delta_c}_{\text{Max}}$ and $\delta$, cf also Tab. \ref{table: table1}), obtaining these values for $U_c$ with error bar $ \sim 0.1$  was a complicated task: 
even for $U=U_c+0.5$, the (expected or fitted) gap $\Delta_c$ is still very small due to the exponential function in Eq. \ref{KT_exp_law}.
Note that for $N=3$ and $N=4$, our proposed values of $U_c$ lie in between the ones calculated through Green's function Monte-Carlo in \cite{assaraf1999}, which read $U_c(N=3)\simeq 2.2$
and  $U_c(N=4)\simeq 2.8$ and the ones calculated in \cite{manmana2011}  which are
$U_c(N=3)\simeq 1.1$ and  $U_c(N=4)\simeq 2.1$.
These latter values were not calculated directly from the charge gaps but rather indirectly, from the fidelity susceptibility of the ground states as a function of $U$.
Moreover, our finite values of $U_c$ shown in Eq. \ref{finite_U_c}, are in contrast with the numerical results obtained in \cite{Szirmai_2007}, which argued for an opening of the gap for infinitesimal $U$ $\forall$ N, and not only for N=2.
In fact, as a result of the implementation of the SU(N) symmetry, we believe that our DMRG simulations are a bit more accurate than the ones in \cite{Szirmai_2007} since they benefit from a lower discarded weight $\mathcal{W}_d^{m,L}$:
while in $\cite{Szirmai_2007}$, the value $\mathcal{W}_d^{m,L} \sim 10^{-6}$ for $N=3, L=90$ and $N=4, L=30$ is reported, 
we have between one and three orders of magnitude less for comparable sizes $L$ (depending on the values of U), as shown in Tab. \ref{table: table1}.

 \begin{table*}[h]
    \centering
\begin{tabular}{|c|c|c|c|c|c|c|} 
\hline
&U=1,\,N=3&U=5,\,N=3&U=1,\,N=4&U=5,\,N=4&U=1,\,N=6&U=5,\,N=6\\
\hline
$m_{\text{Max}}$&8000&8000&10000&10000&16000&12000\\
\hline 
$E_0(m_{\text{Max}},L=36)$ &-47.8504524&-21.9643200&-52.030618&-23.862852&-55.08724&-25.319341 \\
\hline 
$\mathcal{W}_d^{m_{\text{Max}},L=36}$&$7.5 \times 10^{-10}$&$2.8 \times 10^{-12}$&$6.5 \times 10^{-8}$&$1.4 \times 10^{-9}$&$1.1 \times 10^{-6}$&$1.8 \times 10^{-7}$\\
\hline 
$E_0(m=+\infty,L=36)$&-47.8504525&-21.9643200&-52.030621&-23.862852&-55.08726&-25.319345\\
\hline 
$E_0(m_{\text{Max}},L=60)$ &-80.347158&-36.9109353&-87.35688&-40.131088&-92.4828&-42.61863 \\
\hline 
$\mathcal{W}_d^{m_{\text{Max}},L=60}$&$2.0 \times 10^{-8}$&$4.9 \times 10^{-11}$&$7.6 \times 10^{-7}$&$1.4 \times 10^{-8}$&$8.6 \times 10^{-6}$&$1.3 \times 10^{-6}$\\
\hline 
$E_0(m=+\infty,L=60)$&-80.347160&-36.9109353&-87.35691&-40.131089&-92.4831&-42.61867\\
\hline 
$E_0(m_{\text{Max}},L=84)$ &-112.847267&-51.8586464&-122.68571&-56.400596&-129.8790&-59.91933 \\
\hline 
$\mathcal{W}_d^{m_{\text{Max}},L=84}$&$9.2 \times 10^{-8}$&$2.3 \times 10^{-10}$&$2.2 \times 10^{-6}$&$4.8 \times 10^{-8}$&$1.9 \times 10^{-5}$&$2.7 \times 10^{-6}$\\
\hline 
$E_0(m=+\infty,L=84)$&-112.847276&-51.8586465&-122.68585&-56.400600&-129.8804&-59.91947\\
\hline
$e_0(m=+\infty,L=+\infty)$&-1.35428&-0.622858&-1.47213&-0.677939&-1.55829&-0.72092\\
\hline
$L^{\Delta_c}_{\text{Max}}$&102&102&100&100&84&84\\
\hline 
$m_1,m_2$ &4000,6000&4000,6000&6000,8000&6000,8000&8000,12000&8000,12000\\
\hline 
$\Delta_c(m_2,L=L^{\Delta_c}_{\text{Max}})$&0.052&1.13&0.045&0.83&0.040&0.642\\
\hline
$\Delta_c(m=+\infty,L=+\infty)$&0.0022&1.11&0.016&0.81&0.0065&0.596\\
\hline
$c(m=m_{\text{Max}},L=48)$& $2.51\vert 2.77\vert3.11 $ & $1.48\vert1.66\vert2.03$ & $ 3.30\vert3.66\vert4.10$ & $2.15\vert2.48\vert2.90 $ &  $5.11\vert5.49\vert6.55 $ &$4.30\vert4.66\vert5.82$ \\
\hline
$c(m=m_{\text{Max}},L=84)$& $2.69\vert2.84\vert3.04 $ & $1.63\vert1.75\vert2.01$ & $ 3.50\vert3.70\vert3.89$ & $2.41\vert2.60\vert2.84 $  &  $5.07\vert5.55\vert5.83 $ &$4.38\vert4.84\vert5.28$ \\
\hline
\end{tabular}
\caption{Finite-size and finite-m total ground state energies $E_0(m,L)$, discarded weight $\mathcal{W}_d^{m,L}$, charge gaps $\Delta_c(m,L)$, central charges $c(m,L)$ and extrapolated gaps $\Delta_c(m=+\infty,L=+\infty)$, ground state total energies  $E_0(m=+\infty,L)$ and energies per site $e_0(m=+\infty,L=+\infty)$ of the SU(N) FHM on the chain with OBC for L=36, 48, 84 and $L=+\infty$ sites calculated through DMRG simulations with the full SU(N) symmetry, for $N=3,4$ and $6$, filling $1/N$ and for $U=1$ and $U=5$. The number of states kept $m=m_1, m_2 $ and $m=m_{\text{Max}}$ and  the maximal number of sites $L=L^{\Delta_c}_{\text{Max}}$ used for the extrapolation are also shown. For the energies, the constant $-LU/2$ was withdrawn for convenience.
For the central charges, we give the three (ordered in ascending order) values $c_{\text{Floor}(N/2)}, \tilde{c}$ and $c_{0}$, introduced in the text to take into account the Friedel oscillations in the fitting of the entanglement entropy $S(x)$ through the Calabrese-Cardy formula (cf Eq. \ref{calabrese_cardy}). See text for additional details.}
\label{table: table1}
\end{table*}

Moreover, from the infinite DMRG part, we have also calculated the ground state energy per site $e_0(m,L)$ approximated as $e_0(m,L)=(E_0(m,L)-E_0(m,L-2N))/(2N)$
and extrapolated at finite $L$ in the limit $m \rightarrow + \infty$ using the discarded weight  $\mathcal{W}_d^{m,L}$ (cf above), and in the thermodynamical limit $L\rightarrow +\infty$ through a quadratic fitting in $1/L$, to obtain $e_0(m=+\infty,L=+\infty)$ shown in Tab. \ref{table: table1}.

After the infinite size DMRG part, and once the desired length of the chain $L$, was reached, we have also performed some sweeps from left to right and from right to left through the finite-size DMRG. 
For fixed $m$, 
we have computed the entanglement entropy $S(x)$ (cf Eq. (\ref{entanglement_rho})) as a function of $x$, the position of the sweep. For critical spin chains with OBC, the entanglement entropy is given by the Calabrese-Cardy formula \cite{calabrese_cardy}:
\begin{align}
\label{calabrese_cardy}
S(x)=\frac{c}{6} \log\Big{[}\frac{2L}{\pi}\sin\Big{(}\frac{\pi x}{L}\Big{)}\Big{]}+K,
\end{align}
where $K$ is a non-universal constant, and c is the central charge of the associated CFT. 

\begin{figure} 
\centerline{\includegraphics[width=1\linewidth]{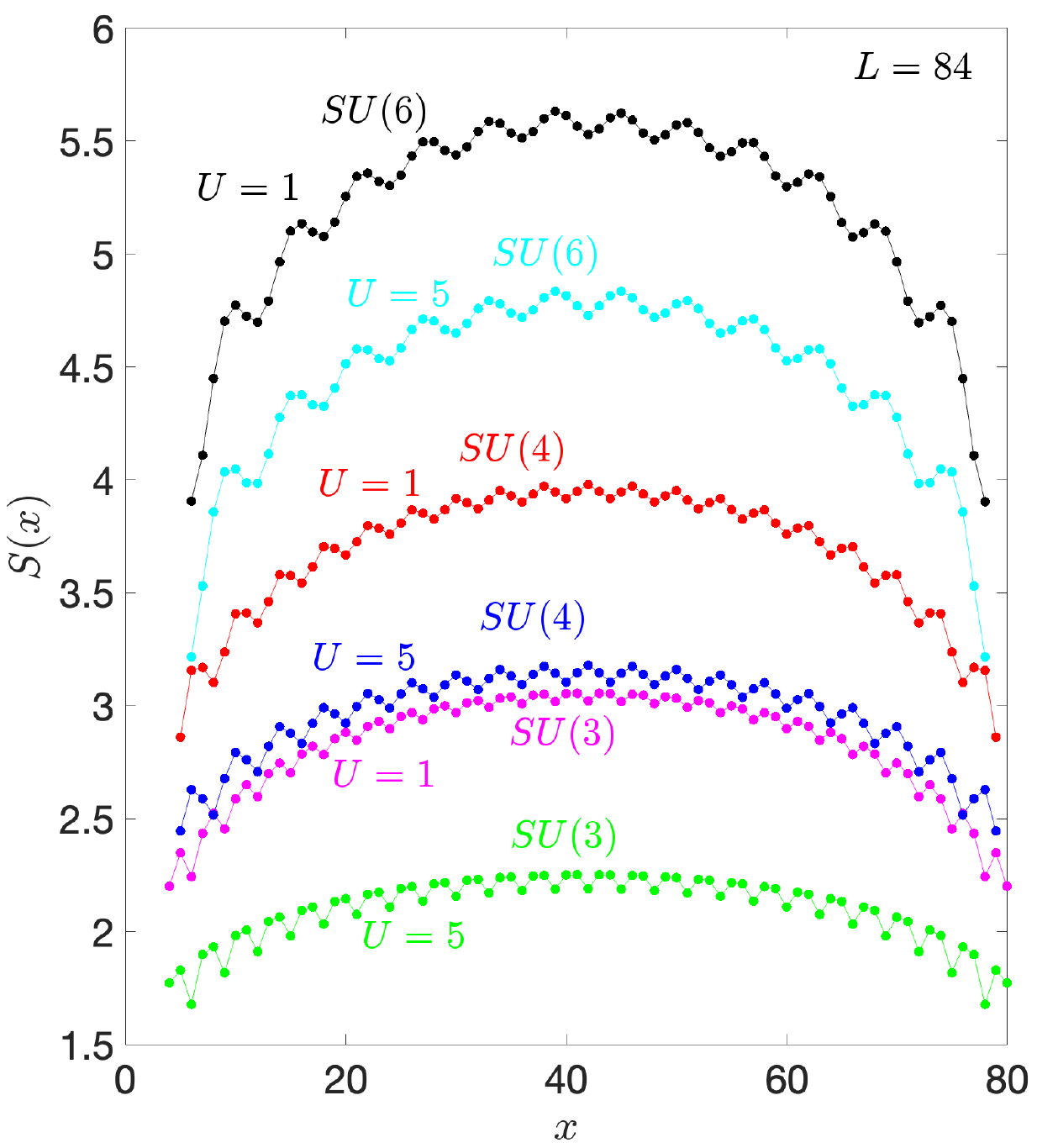}}
\caption{\label{Figure_Entropie} Entanglement entropy $S(x)$ as a function of the position $x$ along an open chain of $L=84$ sites for $N=3,4,6$ and $U=1, 5$ at filling $1/N$. 
The number of states kept for each curve is $m_{\text{Max}}$ shown in Tab. \ref{table: table1}.
Due to the Open Boundary Conditions (OBC), the curves exhibit some Friedel oscillations which are $N$-periodic since the filling is $1/N$. }
\end{figure}

For critical systems, $c$ gives the number of critical modes for the effective  low energy field theory.
According to the bosonization approach developped in \cite{assaraf1999},
it should be here equal to $c_{\text{th}}=N$ for $U< U_c$ and $c_{\text{th}}=N-1$ for $U>U_c$.
In particular, the spin degrees of freedom are described by the SU$(N)_1$  WZW CFT \cite{affleck1986,Affleck_1988} with $N-1$ gapless critical modes for arbitrary $U$,
while the charge degrees of freedom are described by a sine-Gordon model, which becomes critical when $U< U_c$, adding another critical mode, so that $c_{\text{th}}=N-\Theta(U-U_c)$,  where $\Theta$ is the Heaviside step function.
To characterize the two different critical phases, we show in Fig. \ref{Figure_Entropie}  the profile of the entanglement entropy $S(x)$ as a function of the position $x$ for $L=84$,  for $U=1$
, i.e in the  expected metallic phase, and for $U=5$, i.e in the insulating phase for all the values of $N$ considered here, i.e  $N=3,4$ and $6$.

Because of the OBC and of the $1/N$ filling, $S(x)$ has Friedel oscillations with $N$-fold periodicity, that we should remove to fit the central charges.
 We adopt the very same strategy as in \cite{nataf_density_2018}:
 since the oscillations are $N-$periodic, one can plot $S(x)$ as a function of the logarithm of the conformal distance $\frac{1}{6} \log\Big{[}\frac{2L}{\pi}\sin\Big{(}\frac{\pi x}{L}\Big{)}\Big{]}$ separately for different sets of abscissa $x$ of the form $x=N\times p+q$. 
 
 Each set corresponds to a fixed $q=0,1,..N-1$ and to all values of $p$ consistent with the overall length. 
As shown in Fig. \ref{Figure_fitting_entropy} in Appendix \ref{fitting_example}, it then gives rise to several different straight lines (at most $N$, i.e one for each $q$), with different slopes $c_q$. It turns out that one always has $\text{Min}(c_{\text{Floor}(N/2)},c_{0}) \leq  \cdots \leq \text{Max}(c_{\text{Floor}(N/2)},c_{0})$, so that $c_{\text{Floor}(N/2)}$ and $c_{0}$ can be considered as the boundary values of the central charges. 

Alternatively, one can also follow the strategy introduced in \cite{Laflorencie_2006} and developed in \cite{Capponi_oscillations_2013}: Since the Friedel oscillations originate from the bond modulations, it is convenient to fit $\tilde{S}_k(x)=S(x)+k \langle E_{x,x+1} + \text{h.c}\rangle$,
where $\langle E_{x,x+1}\rangle$ is the expectation value on the ground state of the hopping between the left and the right block, and where $k$ is a parameter that is adjusted to best remove the oscillations. From the fitting of $\tilde{S}_k(x)$ with the logarithm of the conformal distance, we extract $\tilde{c}$ (cf Fig. \ref{Figure_fitting_entropy} in Appendix \ref{fitting_example}) and
 we have systematically observed that $\text{Min}(c_{\text{Floor}(N/2)},c_{0}) \leq \tilde{c} \leq \text{Max}(c_{\text{Floor}(N/2)},c_{0})$.

We show in Tab. \ref{table: table1}, these quantities for $L=48$ and $L=84$, and it exhibits good agreement with the field theory expectations, and suggest good evolution with the system sizes: 
$\tilde{c} (L=48),\leq \tilde{c}(L=84) \rightarrow c_{\text{th}}$ and the range given by $\vert c_0-c_{\text{Floor}(N/2)} \vert $ also contracts.
Like for the energies, the convergence of the entanglement entropy is easier in the Mott phase; actually the central charges were already observed to be close to $N-1$ in previous studies, either for the pure Heisenberg SU(N) models \cite{fuhringer2008,nataf_density_2018}, either in the SU(N) FHM for large $U$, (i.e $U=10 t$ in \cite{Szirmai_2008}), although on smaller chains.

In any case, our values of the central charges at $U=1$, indicative of $N$ critical modes, provide further evidence of the presence of a finite metallic phase with $U_c \geq 1$.
Note that in \cite{Szirmai_2007}, for $N=3$, they also observe an half block entropy (directly proportional to $c$, cf Eq. \ref{calabrese_cardy} with $x=L/2$) larger for $U=1$ than for $U=5$.
But they did not interpret the transition between the two phases of different central charges as the metal-insulator transition, while the loss of one critical mode from metallic to insulating phase was already observed numerically for $L=27$ and $N=3$ in \cite{assaraf1999}. They rather considered $U_c$ as the maximum of the block entropy, in contrast with the physical picture built in \cite{assaraf1999}
and validated by our data.

Finally, the form of the finite-size corrections to the central charges, which, for SU(2), is positive and scales as $1/\text{log}(L)^3$ for PBC\cite{affleck_gepner,ziman1987} and is negative and scales as $1/\text{log}(L)^2$ for OBC\cite{hamer,Affleck_Qin_1999}, is
also beyond the scope of our manuscript and would deserve further investigations.

 \section{Conclusions and Perspectives}
To conclude, we have shown how to implement the full SU(N) symmetry in a DMRG code "\`a la White" for the SU(N) FHM using the basis of SSYT.
In particular, we provided many technical details on the calculation of the hopping between the left and the right block, for which the subduction coefficients of the unitary groups play a key role, enabling us to bypass the calculation and the storage of the CGCs. Such a methodology becomes more and more advantageous as $N$ increases (as shown at the end of the section \ref{subduction_coeff}). 
It renders possible the DMRG simulations of the SU(N) FHM on the one dimensional chain at filling $1/N+ \delta/NL$, ($\delta=0,\pm1$), up to $N=6$ and $L=84$ (before extrapolation) with a discarded weight $\mathcal{W}_d^{m,L}$ low enough to get $5$ to $6$ digits ground state energies.
As an application, we computed $U_c$ for the Metal-Insulator transition, 
and we have demonstrated that it is finite for $N>2$, in contrast with some previous DMRG works, not implementing the full SU(N) symmetry \cite{Szirmai_2007}.
Thus, the implementation of non-Abelian symmetries is not merely aimed at slightly refining a numerical result but can also help solve a physical problem and settle a controversy.
We have also calculated the central charges in both the metallic and the insulating phase, confirming the analytical quantum field theory predictions. 
In particular, our numerical results corroborate the presence of $N-\Theta(U-U_c)$, (where $\Theta$ is the Heaviside step function) gapless critical modes, endorsing the spin-charge separation picture elaborated from
the bosonization approach \cite{assaraf1999}: $N-1$ critical modes in the spin sector for every positive $U$, with an additional critical mode in the charge sector when the charge gap vanishes (i.e when $U<U_c$, in the metallic phase).
Our simulations also provide us with accurate ED or DMRG ground state energy values for different parameters, which might be useful for benchmarking future numerical works.

As obvious perspectives, one could first think of addressing different boundary conditions (like the PBC to obtain the structure factors \cite{Mikkelsen_2023} and the Luttinger parameters \cite{manmana2011}, possibly with an additional flux \cite{Chetcuti_2022,Chetcuti_2023}), different fillings \cite{Assaraf_2004,Zhao_2006,Szirmai_2007,Szirmai_2008,nonne2011,Wang_2014,Xu_2018}, other quasi one-dimensional geometries like the ladder\cite{lecheminant2015,Weichselbaum_2018,Capponi_2020}, or models with longer range\cite{Quella_2014} or non-uniform hopping, like the modulated SU(N) FHM \cite{Capponi_2025}.
 
 As a methodological outlook, a useful work would be the extension of the use of the subduction coefficients for the unitary group to other SU(N) symmetry-resolved algorithms based on MPS and to other kind of tensor networks which are available through on-line libraries\cite{Bruognolo_2021,QSpace_1,Vanderstraeten_2019,Fishman_2022}, which are not always adapted to large $N$, as they are based on the calculation/storage/manipulation of CGCs which become computationally prohibitive when $N\geq 6$.

An other important direction of study would be the attempt to implement the SU(N) symmetry in the numerical simulation of the multi-orbital FHM\cite{gorshkov_two_2010,Kobayashi_2012,Bois_2015}.  
In fact, the multi-orbital SU(N) FHM can be regarded as an experimentally feasible class of models that, in the asymptotic limit of large on-site interactions, reduce to SU(N) Heisenberg models with multi-column irreducible representations at each site \cite{Katsura_2008,rachel2009,Bois_2015,dufourPRB2015,PRBNataf2016,Wan2017,Lajko_2017,Quella_2018,Papoular_2019,Gozel_2019,gozel_2020,Herviou_2024}. These spin models have recently attracted significant theoretical interest, as they may host various exotic one-dimensional phases, such as symmetry-protected topological phases \cite{Gu2009,Chen_2010,Fidkowski_2011,Morimoto2014,nonne2013,Quella2013_string,Bois_2015}.
Realizing these models experimentally with cold atoms would require investigating their practically implementable counterpart: the multi-orbital SU(N) FHM.  
 
 Finally, our algebraic approach based on the representations of the unitary group could also be used to study two-dimensional SU(N) FHM which could potentially host SU(N) chiral spin liquids \cite{hermele2009,natafPRL2016,Boos_2020,Chen_2021}.

Work is currently in progress along these lines.
 
 \section{Acknowledgements} 
We thank P. Lecheminant and S. Capponi for valuable discussions on the physical applications of our algorithm and for their critical reading of the manuscript, as well as L. Herviou and F. Mila for their careful reading of the manuscript.
We acknowledge the IT Systems Engineers J. Michel and J-D. Dubois for their instrumental assistance.
The author is supported by the IRP EXQMS project from CNRS.
 \section{Appendix}
 \subsubsection{Basis of semi-standard Young tableaux and matrix elements of the generators of the unitary group $U(L)$ in the Gelfand-Tsetlin representation.}
 \label{GT_rules_appendix}
 The set of semi-standard Young tableaux (SSYT) of a given shape or  Young Diagram (YD) $\alpha=[\alpha_1,\alpha_2,\cdots,\alpha_{L-1},\alpha_L]$ 
 form a basis of the  $U(L)$ irreducible representation (irrep) labelled by this shape.
 A SSYT is filled up with numbers from $1$ to  $L$ in non-descending order from left to right in rows and in (strictly) ascending order from top to bottom in columns, importantly repetition is allowed in rows only.  For instance,
$$
 \ytableausetup{smalltableaux}
\raisebox{1.8ex}{$\ytableaushort{1 3 3 4, 2 4 4, 3 5}$}
$$
is a proper SSYT and defines a basis state of the irrep $[4,3,2,0,0]$ of $U(L\geq5)$. 
Note that the irreps of  U(L) or SU(L)  are basically the same, and the differences occur when considering diagonal generators with or without  vanishing traces.
The SSYTs of a given shape $\alpha=[\alpha_1,\alpha_2,\cdots,\alpha_{L-1},\alpha_L]$ are fully equivalent to the Gelfand-Tstelin patterns \cite{alex2011} of top row $[\alpha_1,\alpha_2,\cdots,\alpha_{L-1},\alpha_L]$.  
From a shape  $\alpha=[\alpha_1,\alpha_2,\cdots,\alpha_{L-1},\alpha_L]$, it is possible to directly calculate $d^{\alpha}_L$, i.e. the number of distinct SSYTs of this shape, which is also the dimension of the irrep $\alpha$ of $U(L)$, through \cite{alex2011}:
\begin{equation}
\label{equation_dim}
d^{\alpha}_L = \prod_{1 \leq j < j' \leq L} \Big{(} 1+\frac{\alpha_j-\alpha_{j'}}{j'-j} \Big{)}.
\end{equation}
It is also possible to introduce an order among the SSYT of a given shape \cite{alex2011}, and the highest weight state (hws) of a given shape $\alpha$, written $\vert \text{HWS} \rangle_{\alpha}$, has its entries in the first row equal to one,
its entries in the second row equal to 2, etc... For instance,
 $$
 \vert \text{HWS} \rangle_{[4 3 2 0 0]} \equiv  \ytableausetup{smalltableaux}
\raisebox{1.8ex}{$\ytableaushort{1 1 1 1, 2 2 2, 3 3}$}
$$
is the hws of the irrep $[4 3 2 0 0]$.

The matrix elements of the infinitesimal generators between equal or consecutive sites $E_{p, p}$, $E_{p-1, p}$ and $E_{p, p-1}$, take simple form derived from group theory results due to Gelfand and Tsetlin \cite{Gelfand_1950}. Calling $\vert \nu \rangle$ a SSYT, one has  for $p=1\cdots L$:
\begin{align}
E_{p, p} \vert \nu \rangle &=\big{(} \# p \in \nu \big{)}\vert \nu \rangle,
\end{align}
where $(\# p \in \nu)$ is equal to the number of occurrences of $p$ inside $\vert \nu \rangle$, corresponding thus to the occupation number on each site [see examples in Fig.~\ref{H_NsNsplus1_chain}]. 
And we have also, for $p=2\cdots L$:
\begin{align}
E_{p-1, p} \vert \nu \rangle & = \sum_{j=1}^{p-1} a^j_{p-1} \vert \nu^{+j}_{p-1}\rangle, \\
E_{p, p-1} \vert \nu \rangle &=\sum_{j=1}^{p-1} b^j_{p-1} \vert \nu^{-j}_{p-1}\rangle, 
\end{align}
where $\vert \nu^{+j}_{p-1}\rangle$ (resp. $\vert \nu^{-j}_{p-1}\rangle$)
is the same SSYT as $\vert \nu \rangle$, except that we have transformed $p$ into $p-1$ (resp. $p-1$ into $p$) in the $j^{th}$  row in $\vert \nu \rangle$. As for the coefficients  $a^j_{p-1}$ and $b^j_{p-1}$, which vanish in case such  transformations are not possible either because there is no $p$ (resp. $p-1$) in the $j^{th}$ row of $\vert \nu \rangle$,
either because the resulting tableau is not a proper SSYT, they read \cite{Vilenkin_vol3}:
\begin{align}
a^j_{p-1} & =  \left | \frac{\prod_{i=1}^p (l_{i,p}-l_{j,p-1})\prod_{i=1}^{p-2} (l_{i,p-2}-l_{j,p-1}-1)}{ \prod_{i\neq j} (l_{i,p-1}-l_{j,p-1})\prod_{i\neq j} (l_{i,p-1}-l_{j,p-1}-1)} \right | ^{1/2}, \label{coefficiens_unitary_generators1} \\
b^j_{p-1} & =  \left | \frac{\prod_{i=1}^p (l_{i,p}-l_{j,p-1} + 1 )\prod_{i=1}^{p-2} (l_{i,p-2}-l_{j,p-1})}{ \prod_{i\neq j} (l_{i,p-1}-l_{j,p-1})\prod_{i\neq j} (l_{i,p-1}-l_{j,p-1}+1)} \right | ^{1/2}, \label{coefficiens_unitary_generators2}
\end{align}
where $l_{k,q}=m_{k,q}-k$ with $m_{k,q}$ the length of the $k^{th}$ row of the sub-tableau that remains when we delete all the boxes containing numbers $>q$ in $\vert \nu \rangle$. 
For instance, one has:
\begin{align}
\label{eq_ssYT}
 \ytableausetup{smalltableaux}
 E_{2,3}\,\raisebox{1.8ex}{$\ytableaushort{1123,2334,456,5}$}=\sqrt{\frac{5}{6}}\,\raisebox{1.8ex}{$\ytableaushort{1122,2334,456,5}$}+\sqrt{\frac{16}{6}}\,\raisebox{1.8ex}{$\ytableaushort{1123,2234,456,5}$},
\end{align}
since,
\begin{align}
l_{1,3}&=3 \hspace{1cm} l_{1,2}=2  \hspace{1cm} l_{1,1}=1 \nonumber \\
l_{2,3}&=1 \hspace{1cm} l_{2,2}=-1  \hspace{.7cm} l_{3,3}=-3,
\end{align}
for  the SSYT in left hand side of Eq. \ref{eq_ssYT}.

Moreover, from successive applications of the commutation relations of the Lie algebra of $U(L)$, 
\begin{equation}
\label{commutation}
[E_{i, j},E_{k, l}]=\delta_{jk}E_{i, l}-\delta_{li}E_{k, j}, \,\,\,\forall 1 \leq i,j \leq L,
\end{equation}

 the generators $E_{p, p+j} \, (j>1)$ are deduced from the generators $E_{p, p+q}$ ($q<j$). We also have $E_{i, j}=E_{j, i}^{\dag}$, so that all the matrices representing  the $E_{i,j}$ operators (for $1 \leq i,j \leq L$) can be obtained easily on the basis of SSYTs.
 
 Finally,  $\vert \text{HWS} \rangle_{\alpha}$  is the only state of shape $\alpha$ satisfying the defining properties of the hws \cite{Paldus_2021}:
\begin{align}
E_{i, i} \vert \text{HWS} \rangle_{\alpha} &=\alpha_i \vert \text{HWS} \rangle_{\alpha}  \,\, \,\,\text{for}\,\,\,\,  1 \leq i\leq L,\label{HWS_eq1} \\
E_{i, j} \vert \text{HWS} \rangle_{\alpha}&=0 \,\, \,\, \text{for}  \,\, \,\,  1 \leq i<j\leq L,\label{HWS_eq2}
 \end{align}
 where the $E_{i, j}$ for $1 \leq i<j\leq L$ are the {\it raising} operators.
 
 \subsubsection{Tensor products of SU(N) irreps and computational cost of calculating Clebsch-Gordan coefficients.}
 \label{tensor_product_appendix}
 Realizing tensor products of SU(N) irreps is needed both in the growing process (cf section \ref{selection_appendix}), and in the construction of the superblock (cf section \ref{reduced_matrix_element}).
 To perform such a tensor product, one can, for instance, use  the Itzykson-Nauenberg  rules \cite{itzykson} (also known as Littlewood Richardson rules \cite{alex2011}),
 we review below. 
 
 Choose one of the two irreps that you want to perform the tensor product of, as the "trunk" (the left one in the following Eq. \ref{eq_tensor_product}), and label the boxes in the first row of the second tableau with "a",
the boxes in the second row with "b", etc...,:
\begin{align}
\label{eq_tensor_product}
 \ytableausetup{smalltableaux}
\raisebox{1.8ex}{$\ytableaushort{\,\,\,\,,\,\,\,,\,\,,}$}\otimes \raisebox{1.8ex}{$\ytableaushort{aaaaaa,bbb,cc,\cdots}$}.
\end{align}
Add one box labelled "a" on the trunk in all possible ways such that it remains a YD (length of rows in non increasing order from top to bottom).
Then, add a second box labelled "a" (if any) requiring that the resultant object is still a YD, etc...
When the boxes labelled "a" are exhausted in the second tableau, add the boxes  labelled "b", etc...
In this process, satisfy the following rules:\\
(i)Keep only the YD with no more than N rows.\\
(ii)Never let two boxes with the same label stand in the same column.\\
(iii)Reading from right to left and top to bottom a resulting tableau, collect the labels of the boxes.\\
One should always find a number of "a"s greater or equal to the number of "b"s, which itself should be greater or equal to the number of "c"s, and so on.\\
(iv)Tableaux with the same attached labels at the  same place should be counted as one. That is to say, two identical representations in the resulting tensorial product of the two shapes should differ by the disposition of the letters.\\
Applying these rules leads for instance to this SU(3) example:
\begin{align}
\label{eq_tensor_product2}
 \ytableausetup{smalltableaux}
\ytableaushort{\,\,,\,}\otimes\ytableaushort{\,\,,\,}&=\bullet \oplus 2 \ytableaushort{\,\,,\,}\oplus  \ytableaushort{\,\,\,} \oplus \ytableaushort{\,\,\,\,,\,\,} \oplus \ytableaushort{\,\,\,,\,\,\,},
\end{align}
where we have deleted N=3-boxes columns and where the outer multiplicity of the shape $[2 1]$ in $[21] \otimes [21]$ is two, while it is one for other irreps.

We now justify that the computational complexity of determining the Clebsch-Gordan coefficients (CGCs), which are necessary for obtaining certain group-theoretical coefficients (such as Wigner 6j and 9j symbols or F-symbols \cite{weichselbaum2020,mcculloch2002,Sahinoglu_2021,Levin_2005}) used in alternative implementations of SU(N) symmetry, scales polynomially with the dimensions of the SU(N) irreps involved in the corresponding tensor product.

We call $C^{\nu,r}_{\nu_1,\nu_2}$, the CGCs of  $\vert \nu \rangle_r $ with respect to $\vert \nu_1 \rangle $ and $\vert \nu_2 \rangle $. 
$\vert \nu \rangle_r$ is a SSYT of shape $\alpha$, which labels an SU(N) irrep which appears in the tensor product of $\alpha_1$ and $\alpha_2$, and $\vert \nu_i \rangle$ is a SSYT of shape $\alpha_i$ ($i=1,2$). $r$ is the integer index that accounts for possible outer multiplicity, (i.e. $r=1,2, \cdots T^{\alpha}(\alpha_1,\alpha_2)$, where $T^{\alpha}(\alpha_1,\alpha_2)$ is the outer multiplicity of $\alpha$ into $\alpha_1 \otimes \alpha_2$).
The basis states of the irrep  $\alpha$ are then written in terms of CGCs through:
\begin{align}
\label{Clebsch_eq}
\vert \nu \rangle_r= \sum_{\nu_1,\nu_2} C^{\nu,r}_{\nu_1,\nu_2} \vert \nu_1 \rangle \otimes  \vert \nu_2 \rangle.
\end{align} 

As well explained in \cite{alex2011}, the ansatz to determine the CGCs $C^{\nu,r}_{\nu_1,\nu_2}$ is based on the application of the generator operators
$E_{i, j}\equiv E_{i, j}^{\alpha_1} \otimes   \mathcal{I} ^{\alpha_2}+ \mathcal{I} ^{\alpha_1} \otimes  E_{i, j}^{\alpha_2}$, where the $ E_{i, j}^{\alpha_i}$ (resp. $ \mathcal{I} ^{\alpha_i}$) are the matrices of the SU(N) generators (resp. the identity) on the irrep $\alpha_i$ (for $i=1,2$), on both side of the Eq. (\ref{Clebsch_eq}), and on the use of the known coefficients of the matrix representation of these operators (cf Eq. (\ref{coefficiens_unitary_generators1}) and (\ref{coefficiens_unitary_generators2})).

 One starts with the hws for $\alpha$: $\vert \nu \rangle_r \rightarrow \vert \text{HWS} \rangle_{\alpha}$. We apply first Eq. (\ref{HWS_eq1}) to obtain some selection rules: some $C^{\nu,r}_{\nu_1,\nu_2}$ are trivially 0 when there is not the same total number of $p=1 \cdots N$ in the SSYTs on both sides of Eq. (\ref{Clebsch_eq}). 
Then Eq. (\ref{HWS_eq2}) applied on Eq. (\ref{Clebsch_eq}) leads to a linear system of equations in the CGCs, with possibly several solutions in the case of outer multiplicity $T^{\alpha}(\alpha_1,\alpha_2)>1$, which implies some ambiguity in the CGCs that is acceptable or not depending on the applications.
Then, as for SU(2), to obtain the CGCs for the other $\vert \nu \rangle_r$, we apply the {\it lowering} operators $E_{i, j}$ (for $L \geq i>j\geq 1$) on both side of  Eq. (\ref{Clebsch_eq}) which creates some linear combination over several states $\vert \nu \rangle_r$, so that one must determine the CGCs of all basis states.
The solution of the successive systems of equations (which might include the reduction of the ambiguity caused by $T^{\alpha}(\alpha_1,\alpha_2)>1$) requires some algebraic manipulation whose complexity scales polynomially in both the dimensions of the irreps $\alpha_1$ and $\alpha_2$.

 \subsubsection{Selection of states for the current step and construction of the new matrices for the left block}
\label{selection_appendix}
We explain here how to pass from the stage where both the {\it left} and the {\it right} blocks have $N_s$ sites to the stage where they both have $N_s+1$ sites.
We focus below on the modifications required for studying the SU(N) FHM compared to the SU(N) Heisenberg model \cite{nataf_density_2018} (mainly due to the occupation number which is not fixed any more on each site). Thus, a quick review the previous article \cite{nataf_density_2018} may also be useful, although the current section is self-contained.

Focusing on the left block (cf Fig. \ref{sketch_systeme}),  we assume that we have kept in memory $m_{N_s}\leq m$ states from the previous stage, each of them belonging to a given SU(N) symmetry sector $\bar{\alpha}$ (seen as an $U(N_s)$ irrep).
\begin{align}
\label{states_alpha}
\{\vert\zeta^{\bar{\alpha}}_1\rangle,\vert\zeta^{\bar{\alpha}}_2\rangle,\cdots,\vert\zeta^{\bar{\alpha}}_{m^{\bar{\alpha}}_{N_s}}\rangle\}.
\end{align}
The number of states $m^{\bar{\alpha}}_{N_s}$ satisfy:
\begin{align}
\sum_{\bar{\alpha}} m^{\bar{\alpha}}_{N_s}=m_{N_s},
\end{align}
where the sum runs over the shapes $\bar{\alpha}$ fulfilling the constraints for the number of rows/columns/boxes explained in the main text. 

The states in Eq. (\ref{states_alpha}) are the eigenstates of the $m^{\bar{\alpha}}_{N_s} \times m^{\bar{\alpha}}_{N_s}$ reduced density matrix $\rho^{\bar{\alpha}}$ (we will show in Appendix \ref{superblock_appendix} how to calculate the reduced density matrix for a given sector $\bar{\alpha}$). The corresponding positive eigenvalues are the {\it DMRG weights} and are ranked from the largest to the lowest: $\{\lambda^{\bar{\alpha}}_1,\lambda^{\bar{\alpha}}_2,\cdots,\lambda^{\bar{\alpha}}_{m^{\bar{\alpha}}_{N_s}}\}$. In addition, we also assume that from the previous stage we have kept the matrices $\mathcal{H}^{\bar{\alpha}}_{N_s}$, which are the matrices of the FHM $H_{N_s}$ with OBC for $N_s$ sites, expressed in the basis of Eq. (\ref{states_alpha}): the $(i,j)$ coefficient of $\mathcal{H}^{\bar{\alpha}}_{N_s}$ is $(\mathcal{H}^{\bar{\alpha}}_{N_s})_{i,j}=\langle \zeta^{\bar{\alpha}}_i\vert H_{N_s}\vert\zeta^{\bar{\alpha}}_j \rangle$.

\begin{figure} 
\centerline{\includegraphics[width=.8\linewidth]{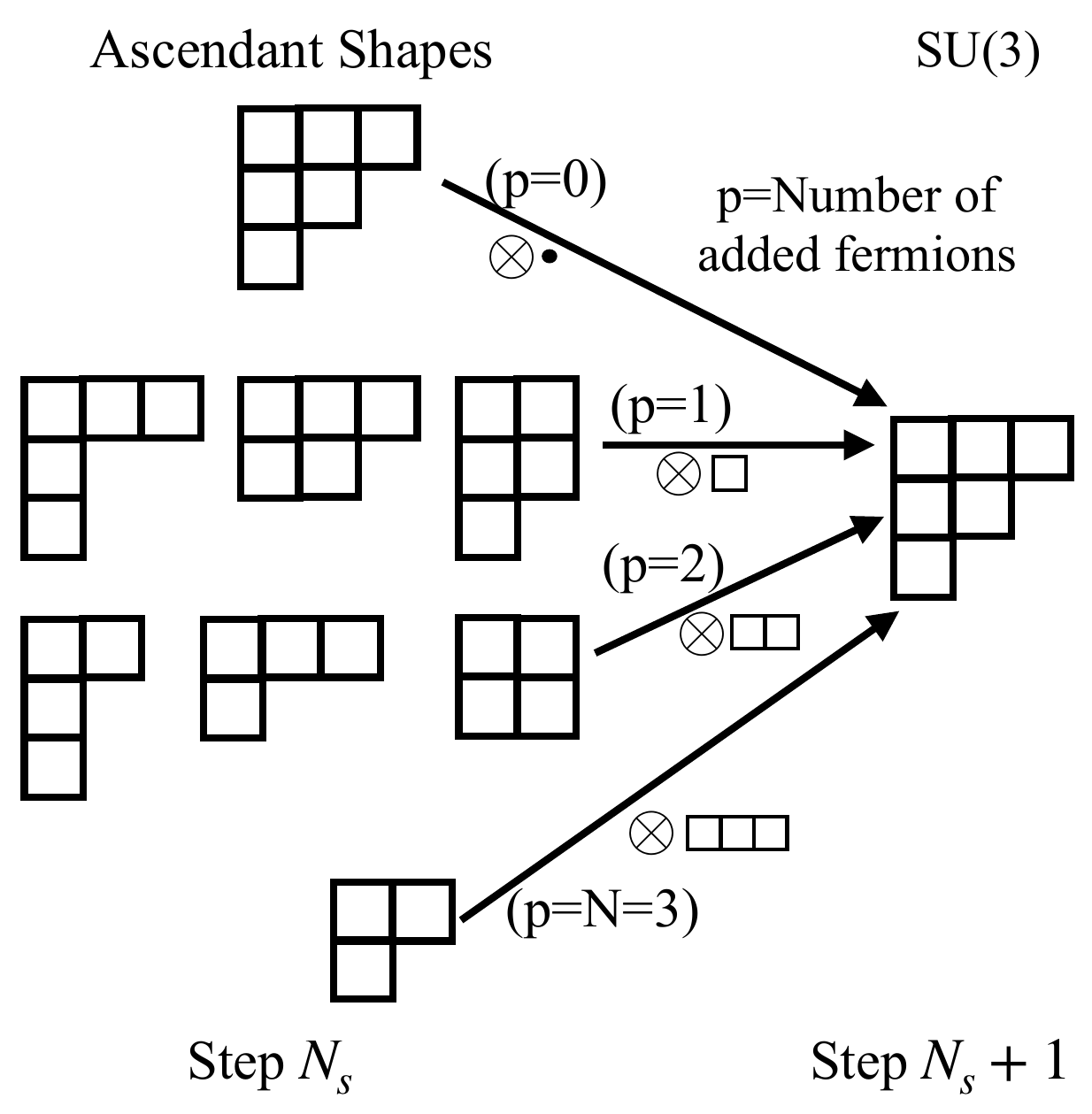}}
\caption{\label{sketch_irreps_ascendant} 
We show here examples of {\it ascendant} shapes $\bar{\alpha}$ (on the left) for the shape $\bar{\beta}=[3 2 1]$ for $N=3$.
$\bar{\beta}$ should belong to $\bar{\alpha} \otimes [p]$ ($p=0,1,\cdots, N$), where $[p]$ is the one-row fully symmetric irrep with $p$ boxes.
See text for details.
}
\end{figure}

To add a new site and select $m_{N_s+1} \leq m$ states, we scrutinize every shape $\bar{\beta}$ which satisfies the constraints (i.e less than $N_s+1$ rows, $N$ columns and $f \times N \times (2N_s+2)$ boxes), and is equivalent (after transposition) to one of the $M$  input SU(N) irreps. There are $M_{N_s+1}$ such shapes. For each shape $\bar{\beta}$, we consider all the possible {\it ascendant} shapes $\bar{\alpha}$: they are such that $\bar{\beta}$ belongs to the tensor product $\bar{\alpha} \otimes [p]$ for $p=0,1,\cdots N$, where $[p]$ is the one-row fully symmetric irrep of $p$ boxes, which is the transpose of the p-box single column irrep, which is the irrep of the added single site with $p$ fermions on it (See Fig. \ref{sketch_irreps_ascendant} for an SU(3) example). 

Like in \cite{nataf_density_2018}, for every shape $\bar{\beta}$, we create $L_{\bar{\beta}}$, the list containing all the DMRG weights $\lambda^{\bar{\alpha}}_q$ of  all the associated ascendant shapes $\bar{\alpha}$, and we create then $L_{N_s+1}$, the union of these lists $L_{N_s+1}=\underset{\bar{\beta}}{\cup}L_{\bar{\beta}}$. We choose the $m_{Ns+1}$  largest values in $L_{N_s+1}$, where $m_{Ns+1}=\text{Min}(m,\text{cardinal}(L_{N_s+1}))$, and the labels attached to each of the chosen $\lambda^{\bar{\alpha}}_q$  allows one to select for each sector $\bar{\beta}$ and each ascendant shape $\bar{\alpha}$,  $m^{\bar{\alpha}}_{\bar{\beta}, N_s+1}$ ascendant states, so that  $m^{\bar{\beta}}_{N_s+1}=\sum_{\bar{\alpha}}m^{\bar{\alpha}}_{\bar{\beta}, N_s+1}$ will be the dimension of the subspace corresponding
to the irrep $\bar{\beta}$ in the left (or right) block of size $N_s+1$.
Calling $\sigma_{\beta}$ the sum of the corresponding eigenvalues, the weight discarded by the current selection (truncation) is 
\begin{equation}
\mathcal{W}_d^{m,L}=1-\sum_{\beta} g_{\beta} \sigma_{\beta},
\end{equation}
where $g_{\beta}=\text{dim}(\beta)/h$, with $\text{dim}(\beta)$, the dimension of the SU(N) irrep of shape $\beta \equiv \bar{\bar{\beta}}$ (before transposition), and where $h$ is the dimension of the local Hilbert space,
 i.e. $h=2^N$. Note that for the Heisenberg model with the fundamental SU(N) irrep on each site, the same formula applied but with $h=N$ \cite{nataf_density_2018}.

In addition, the set of numbers $\{m^{\bar{\alpha}}_{\bar{\beta}, N_s+1}\}_{\bar{\alpha}}$ (stage $N_s+1$) and $\{m^{\bar{\chi}}_{\bar{\alpha}, N_s}\}_{\bar{\chi}}$ (stage $N_s$), define a genealogy for each state in the sector $\bar{\beta}$ up to the {\it grandparent} level.
These set of numbers, and the set of wave-functions  $\{\vert\zeta^{\bar{\alpha}}_1\rangle,\vert\zeta^{\bar{\alpha}}_2\rangle,\cdots,\vert\zeta^{\bar{\alpha}}_{m^{\bar{\alpha}}_{N_s}}\rangle\}$, $\forall \bar{\alpha}$ an ascendant shape of $\bar{\beta}$, are the two sufficient ingredients, with the GT rules for the coefficients of the generators of $U(N_s+1)$ (cf Appendix \ref{GT_rules_appendix}), to build the new matrices for the left block.
In fact, the FHM Hamiltonian for $N_s+1$ sites, can be decomposed as: 
\begin{align} 
\label{H_Nsplus1}
H_{N_s+1}=H_{N_s}-t(E_{N_s,N_s+1}+h.c)+\frac{U}{2} E_{N_s+1,N_s+1}^2.
\end{align}
First,  the matrix representing $H_{N_s}$ in the sector $\bar{\beta}$ (of size $m^{\bar{\beta}}_{N_s+1}\times m^{\bar{\beta}}_{N_s+1}$) is just the concatenation of the submatrices $(\mathcal{H}^{\bar{\alpha}}_{N_s})_{i,j}$ where $1\leq i \leq m^{\bar{\alpha}}_{\bar{\beta}, N_s+1} ,1\leq j \leq m^{\bar{\alpha}}_{\bar{\beta}, N_s+1}$, and where the matrices $\mathcal{H}^{\bar{\alpha}}_{N_s}$ were kept in memory from the previous stage.

\begin{figure} 
\centerline{\includegraphics[width=\linewidth]{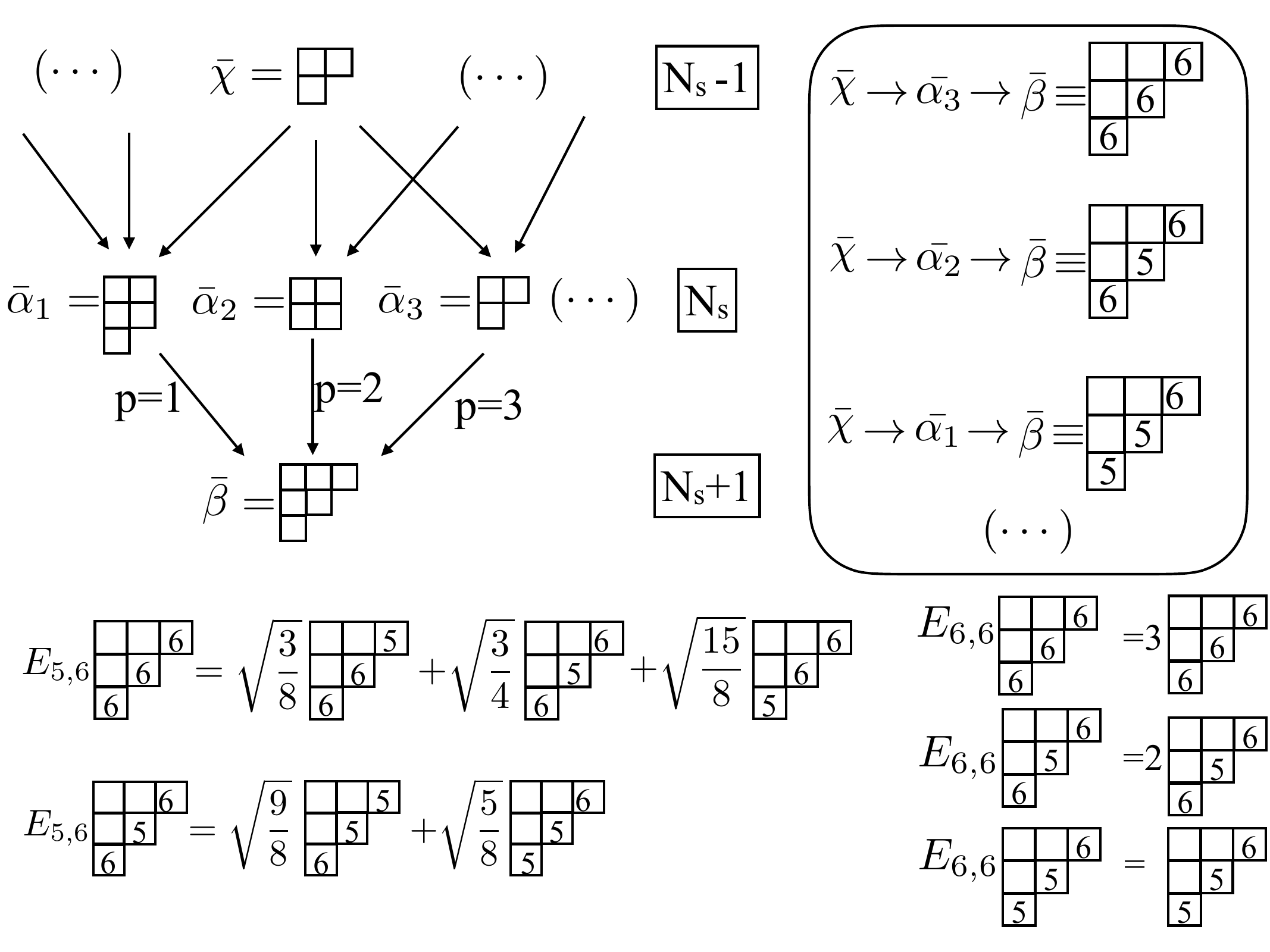}}
\caption{\label{H_NsNsplus1_chain} To create $E_{N_s,N_s+1}+h.c$ and  $E_{N_s+1,N_s+1}$ on the sector labelled by a shape $\bar{\beta}$, one just needs to know
the chain of shapes $\bar{\chi} \rightarrow \bar{\alpha} \rightarrow \bar{\beta}$ which characterize the different class of states, as it determines the locations and the number of occurrences of the numbers $N_s$ and $N_s+1$ on each SSYT ($N_s=5$ here). In particular, $p$ designates here the number of occurences of $N_s+1$ in the SSYT, and also the number of added fermions on the site $N_s+1$.
We apply the GT rules (cf Appendix \ref{GT_rules_appendix}) to get the coefficients of the generators $E_{N_s,N_s+1}$ and $E_{N_s+1,N_s+1}$ on each class of states.  In particular, $E_{N_s,N_s+1}$ or $E_{N_s+1,N_s}$ applied on a SSYT  just acts on the boxes containing $N_s$ and $N_s+1$. See text for details.
}
\end{figure}
Furthermore, the genealogy of each state in the sector $\bar{\beta}$, characterized by the chain of shapes $\bar{\chi} \rightarrow \bar{\alpha} \rightarrow \bar{\beta}$, 
is enough to calculate the matrices representing both $E_{N_s+1,N_s+1}$ and $E_{N_s,N_s+1}$ (and its h.c) on the sector  $\bar{\beta}$, as it gives the locations (and the number of occurrences) of the indices $N_s$ and $N_s+1$ in the SSYTs, which form the underlying basis on which each state is decomposed.
Firstly, as illustrated in Fig. \ref{H_NsNsplus1_chain}, the part of the chain $\bar{\alpha} \rightarrow \bar{\beta}$ tells us  how many boxes one adds from $\bar{\alpha}$ to $\bar{\beta}$, which is nothing but the number of fermions on site $N_s+1$ (equal to the operator $E_{N_s+1,N_s+1}$), also equal to the number of occurrences of the index $N_s+1$ in every SSYT.
Secondly, for a matrix element $\langle \zeta^{\bar{\alpha}}_i\vert E_{N_s,N_s+1} \vert\zeta^{\bar{\alpha}'}_j \rangle$,
not to be zero, the "grand father" shape $\bar{\chi}$ should be common, and the number of occurrences of $N_s+1$ (resp. $N_s$) in the SSYT on which  $\vert\zeta^{\bar{\alpha}'}_j \rangle$ is decomposed should be one more (resp. one less) than such number in $\vert \zeta^{\bar{\alpha}}_i\rangle$.
The Gelfand-Tstelin coefficients for the generators of the unitary group\cite{Gelfand_1950} (cf Appendix \ref{GT_rules_appendix}) illustrated in Fig. \ref{H_NsNsplus1_chain}, and proper overlap of vectors of coefficients (split according to the numbers $\{m^{\bar{\chi}}_{\bar{\alpha}, N_s}\}_{\bar{\chi}}$), enable us to calculate the matrix representing $E_{N_s,N_s+1}$ on the sector $\bar{\beta}$.

\subsubsection{Hilbert space of the superblock and computation of the reduced density matrices}
\label{superblock_appendix}
From the matrix $T^{\bar{\gamma}_L}_{N_s+1}$ (cf section \ref{reduced_matrix_element}), we list the $M^{GS}_{N_S+1} \leq M_{N_S+1}$ {\it relevant} shapes $\tilde{\beta}_k$ (for $1\leq k \leq  M_{N_S+1}$), which are such that the column $T^{\bar{\gamma}_L}_{N_s+1}(:,k)$ have at least one non zero entry. 
The Hilbert space of the superblock on the sector $\bar{\gamma}_L$ is then the direct sum of the tensor product of the sectors corresponding to the shapes $\tilde{\beta}_k$ (for $k=1,\cdots,M^{GS}_{N_S+1}$) for the left block otimes sector labelled by the shapes $\tilde{\beta}_{k'}$  for the right block with a multiplicity equal to $T^{\bar{\gamma}_L}_{N_s+1}(k,k')$.
Consequently, the matrices $\mathcal{H}_{N_s+1}^{\text{Left}}$ and $\mathcal{H}_{N_s+1}^{\text{Right}}$ representing respectively $H_{N_s+1}^{\text{Left}}$ and $H_{N_s+1}^{\text{Right}}$ on the sector $\gamma_L$ of the superblock, are:
\begin{align}
\mathcal{H}_{N_s+1}^{\text{Left}}&=\underset{T^{\bar{\gamma}_L}_{N_s+1}(q,q')>0 }{\bigoplus} \overset{T^{\bar{\gamma}_L}_{N_s+1}(q,q')}{ \underset{r=1}{\bigoplus}}\mathcal{H}^{\tilde{\beta}_q}_{N_s+1} \otimes \mathcal{I}_{N_s+1}^{\tilde{\beta}_{q'}} \nonumber \\
\mathcal{H}_{N_s+1}^{\text{Right}}&=\underset{T^{\bar{\gamma}_L}_{N_s+1}(q',q)>0 }{\bigoplus}  \overset{T^{\bar{\gamma}_L}_{N_s+1}(q',q)}{ \underset{r=1}{\bigoplus}} \mathcal{I}_{N_s+1}^{\tilde{\beta}_{q'}}  \otimes \mathcal{H}^{\tilde{\beta}_q}_{N_s+1},
\end{align}
where $\mathcal{I}_{N_s+1}^{\tilde{\beta}_q}$ is the $m^{\tilde{\beta}_q}_{N_s+1}  \times m^{\tilde{\beta}_q}_{N_s+1} $ identity matrix on the sector labelled by the shape $\tilde{\beta}_q$ (for the left or the right blocks). The basis for the superblock is made of vectors of the form $\vert \zeta^{\tilde{\beta}_q}_i, \zeta^{\tilde{\beta}_{q'}}_j, r \rangle$, for $1 \leq r \leq T^{\bar{\gamma}_L}_{N_s+1}(q,q')$, $1 \leq i \leq m^{\tilde{\beta}_q}_{N_s+1}$ and $1 \leq j \leq m^{\tilde{\beta}_{q'}}_{N_s+1}$, and the dimension of the superblock is thus $  \sum_{q,q'}T^{\bar{\gamma}_L}_{N_s+1}(q,q') \times m^{\tilde{\beta}_q}_{N_s+1} \times m^{\tilde{\beta}_{q'}}_{N_s+1}$.
 Once the matrix for the hopping between the left and the right block, i.e. $E^{\gamma_L}_{N_s+1,N_s+2} + h.c $,  is obtained on the Hilbert space of the superblock (cf section \ref{reduced_matrix_element} and Appendix \ref{subduction_coeff}),   one gets $\mathcal{H}^{\bar{\gamma}_L}_{L}$, i.e the matrix representing $H_L$ on ${\bar{\gamma}}_L$, that we need to diagonalize (using for instance the Lanczos algorithm)  to get the ground state $ G^L_{\bar{\gamma}_L}$. 

Then, calling $V_i^k$ the vector of indices of every states of the form $\vert \zeta^{\tilde{\beta}_k}_i, \cdots, \cdots \rangle$ in the full Hilbert space, for  $k=1,\cdots, M^{GS}_{N_s+1}$ and $1\leq i \leq m^{\tilde{\beta}_k}_{N_s+1}$, one computes the reduced density matrices $\rho^{\tilde{\beta}_k}$, whose coefficients are (for $1\leq i,j \leq m^{\tilde{\beta}_k}_{N_s+1}$) $
\rho^{\tilde{\beta}_k}(i,j)=\text{dim}(\bar{\tilde{\beta}}_k)^{-1} G^{L,\dagger}_{\bar{\gamma}_L}(V_i^k)G^L_{\bar{\gamma}_L}(V_j^k)$,
where $\text{dim}(\bar{\tilde{\beta}}_k)^{-1}$ is the inverse of the dimension of the SU(N) irrep of shape $\bar{\tilde{\beta}}_k$ (which is the shape whose transposition gives $\tilde{\beta}_k$), and which guarantees the correct normalization of the reduced density matrices. 
We then diagonalize the reduced density matrices to obtain the set of eigenvalues ranked from the largest to the lowest one: $\{\lambda^{\tilde{\beta}_k}_1,\lambda^{\tilde{\beta}_k}_2,\cdots,\lambda^{\tilde{\beta}_k}_{m^{{\tilde{\beta}_k}}_{N_s+1}}\}$, as well as the corresponding eigenvectors :
$
\{\vert\zeta^{\tilde{\beta}_k}_1\rangle,\vert\zeta^{\tilde{\beta}_k}_2\rangle,\cdots,\vert\zeta^{\tilde{\beta}_k}_{m^{\tilde{\beta}_k}_{N_s+1}}\rangle\}, 
$
and we perform a rotation to re-express $\mathcal{H}^{\tilde{\beta}_k}_{N_s+1}$ in this basis. We keep in memory those quantities for the next stage.
At this step, one can also calculate the entanglement entropy $S(L)$:
\begin{align}
\label{entanglement_rho}
S(L)=-\sum_{k=1}^{M^{GS}_{N_S+1}} \text{dim}(\bar{\tilde{\beta}}_k) \text{Tr} [\rho^{\tilde{\beta}_k} \log (\rho^{\tilde{\beta}_k})],
\end{align}
where the coefficients $\text{dim}(\bar{\tilde{\beta}}_k) $ account for the multiplicities (cf \cite{Sierra_1996,weichselbaum2012,nataf_density_2018}).
Note that the matrices representing the FHM in the irrelevant sectors, i.e.: 
\begin{align}
\mathcal{H}^{\bar{\beta}}_{N_s+1}\,\,\text{for}\,\, \bar{\beta} \not \in \{\tilde{\beta}_1, \tilde{\beta}_2,\cdots,\tilde{\beta}_{M^{GS}_{N_S+1}}\} \nonumber
\end{align}
 do not undergo any transformation at this step.

 \subsubsection{Calculation of the reduced matrix elements using the $U(m+n)\supset U(m) \otimes U(n)$ subduction coefficients.}
 \label{subduction_coeff}
  In this section, we will show how to calculate the  reduced matrix elements $\langle \bar{\beta}_{q_3},l_3  \vert \otimes\langle \bar{\beta}_{q_4},l_4 \vert  E_{\bar{\gamma}_L} \vert \bar{\beta}_{q_1},l_1\rangle \otimes \vert \bar{\beta}_{q_2},l_2\rangle$, where  $\bar{\beta}_{q_j}$ is a shape with at most $N$ columns (i.e. transposition of a SU(N) irrep or Young diagram), and $l_j$ is the vector of rows (conventionally ranked in descending order) of the cross located at the bottom corners \cite{PRBNataf2016} \footnote{For a given shape $\bar{\beta}$, the bottom corners correspond to all the rows $j$ such that $\bar{\beta}_j > \bar{\beta}_{j+1}$.} inside $\bar{\beta}_{q_j}$, for $j=1,2,3,4$. 
 
  We call $n_j$ the length of the vector $l_j$, i.e the number of cross in the shape $\bar{\beta}_{q_j}$  for $j=1,2,3,4$  .
The idea is to use the SSYTs and the Gelfand-Tsetlin representation of the unitary group (cf Appendix\ref{GT_rules_appendix}) to compute this  kind of coefficient.

Let's first notice that one condition for these coefficients not to vanish is that when one withdraws a box containing a cross in $\bar{\beta}_{q_2}$ to add it to $\bar{\beta}_{q_1}$,
 one should obtain respectively $\bar{\beta}_{q_4}$ and $\bar{\beta}_{q_3}$ with their cross located in the good position.
 Secondly, the value of these coefficients depends on the relative position of the cross for the left block with respect to the cross for the right block in the shape $\bar{\gamma}_L$,
   so that empty $N$-boxes row could be safely deleted/added from the top in $\bar{\beta}_{q_1} $, $ \bar{\beta}_{q_2}$,  $\bar{\beta}_{q_3}$, $ \bar{\beta}_{q_4}$ and $\bar{\gamma}_L$ \footnote{
  Equivalently, the shapes before transposition represent the same SU(N) irreps independently of the empty $N-$boxes column that one could add/withdraw, since we calculate SU(N) group theory coefficients.}.
  Thus, one has for instance:
   \begin{equation}
\text{\raisebox{-2.4ex}{$\mathlarger{\mathlarger{\mathlarger{\mathlarger{\mathlarger{\mathlarger{\langle}}}}}}$}}  \ytableaushort{\,\,\,\,,\,\,\times,\,\times}\otimes   \ytableaushort{\,\,\,\,,\,\,\times,\,,\,}\text{\raisebox{-2.4ex}{$\mathlarger{\mathlarger{\mathlarger{\mathlarger{\mathlarger{\mathlarger{\vert}}}}}}$}}E_{\text{\small $\ytableaushort{\,\,\,\,,\,\,\,\,,\,\,\,\,,\,\,\,,\,\,\,}$}} \text{\raisebox{-2.4ex}{$\mathlarger{\mathlarger{\mathlarger{\mathlarger{\mathlarger{\mathlarger{\vert}}}}}}$}}  \ytableaushort{\,\,\,\,,\,\,\times,\,}\otimes   \ytableaushort{\,\,\,\,,\,\,\times,\,\times,\,}\text{\raisebox{-2.4ex}{$\mathlarger{\mathlarger{\mathlarger{\mathlarger{\mathlarger{\mathlarger{\rangle}}}}}}$}} \nonumber
\end{equation} 
\begin{equation}
\text{\raisebox{-1.4ex}{$=$}}
\text{\raisebox{-2.4ex}{$\mathlarger{\mathlarger{\mathlarger{\mathlarger{\mathlarger{\mathlarger{\langle}}}}}}$}}  \ytableaushort{\,\,\times,\,\times}\otimes   \ytableaushort{\,\,\times,\,,\,}\text{\raisebox{-2.4ex}{$\mathlarger{\mathlarger{\mathlarger{\mathlarger{\mathlarger{\mathlarger{\vert}}}}}}$}}E_{\text{\small $\ytableaushort{\,\,\,\,,\,\,\,,\,\,\,}$}} \text{\raisebox{-2.4ex}{$\mathlarger{\mathlarger{\mathlarger{\mathlarger{\mathlarger{\mathlarger{\vert}}}}}}$}}  \ytableaushort{\,\,\times,\,}\otimes   \ytableaushort{\,\,\times,\,\times,\,}\text{\raisebox{-2.4ex}{$\mathlarger{\mathlarger{\mathlarger{\mathlarger{\mathlarger{\mathlarger{\rangle}}}}}}$}}
\end{equation}

Consequently, the number of such coefficients is finite when one considers only the  $M$ first SU(N) irreps.
An upper boundary for this number of coefficients is $M \times 2^N \times 2^N \times N^2$.
In fact, for a given shape, there are at most $\binom{N}{k}$ ways to locate $k$ cross, for $k=0,1,\cdots N$, which gives a first factor $2^N$.
For each of these shape+cross, one has one partner shape such that the tensor product might give $\bar{\gamma}_L$, with at most $2^N$ different configurations for the cross.
Finally, given such a {\it ket}, one has at most $N^2$ ways to delete one cross in the right block shape to add it in the left block shape.
Such a number is an overestimation by a factor $N$ typically, so for $M=300$, depending on $\bar{\gamma}_L$, one has typically $\sim 10^7$ non zero coefficients 
for $N=6$ and $ \sim 10^5$ non zero  coefficients for $N=3$.

Let's now give the general methodology focusing on the example provided in Eq. \eqref{coeff_ex}.

  \underline{Step 1}

- We first replace the couple $(\bar{\beta}_{q_1},l_1)$ by a SSYT $S_1$ of the same shape, having its last numbers (equal to $r_1+1$, where $r_1$ is the number of rows of the shape $\bar{\beta}_{q_1}$ without the cross) located in the bottom corners $l_1$: $(\bar{\beta}_{q_1},l_1)\rightarrow S_1$. For our example, $(\bar{\beta}_{q_1},l_1)=([4,3,1],[1])$, one has $r_1=3$ and:
\begin{align}
 \ytableaushort{\,\,\,\times,\,\,\,,\,} \rightarrow \ytableaushort{1114,222,3}=S_1.
\end{align}  
- For the second couple $(\bar{\beta}_{q_2},l_2)$, we let aside the vector $l_2$ for step 3, and we first create $S_2^{\text{hws}}$ the highest weight state (hws) SSYT of shape  $\bar{\beta}_{q_2}$, and then we reindex the numbers in $S_2^{\text{hws}}$ to get $\tilde{S}_2^{\text{hws}}$: $1 \rightarrow L$,
$2\rightarrow L-1$, etc, where $L=r_1+r_2+2$, where $r_2$ is the number of rows of $\bar{\beta}_{q_2}$ without the boxes containing the cross.
Then, for our example, one has $(\bar{\beta}_{q_2},l_2)=([4,3,2,1],[3 2])$ and $r_2=4$, $L=9$:
 \begin{align}
 \ytableaushort{\,\,\,\,,\,\,\times,\,\times,\,} \rightarrow \ytableaushort{1111,222,33,4}=S_2^{\text{hws}}\rightarrow \ytableaushort{9999,888,77,6}=\tilde{S}_2^{\text{hws}}.
\end{align}    
 \underline{Step 2}
 
 - We then expand the product $S_1 \otimes \tilde{S}_2^{\text{hws}} $ on the shape $\bar{\gamma}_L$ using the $U(r_1+r_2+2)\supset U(r_1+1) \otimes U(r_2+1)$ subduction coefficients \cite{chen}.
 One must find a linear superposition of SSYTs of shape $\bar{\gamma}_L$ having the same properties of "internal symmetries" between particles as the 
product $S_1\otimes S_2^{\text{hws}}$.
In particular, $S_1\otimes \tilde{S}_2^{\text{hws}}$ should have its first  $r_1+1$ entries like in $S_1$, and must satisfy the {\bf defining properties of the hws} (cf Eqs. \eqref{HWS_eq1} and \eqref{HWS_eq2} ) of the shape $\bar{\beta}_{q_2}$ but applied on the numbers $L$, $L-1$,$\cdots$ and $L-r_2+1$.
Thus, as a basis set of our first expansion, we create all the SSYTs of shape $\bar{\gamma}_L$, starting like $S_1$, and with $L$ in the last $\bar{\beta}_{q_2}(1)$ bottoms corners,  with $L-1$ in the following $\bar{\beta}_{q_2}(2)$ bottom corners, etc $\cdots$. In particular, no equal number should appear in the same column.
Thus, for our example, one should expand $S_1\otimes \tilde{S}_2^{\text{hws}}$ on the set:
\begin{align}
 \label{basis}
\begin{ytableau}
 1&1 & 1&4\\2 &2&2&8 \\3&7&7&9\\6 &8&8\\ 9& 9&9
 \end{ytableau}
 \hspace{0.3cm}
 \begin{ytableau}
 1&1 & 1&4\\2 &2&2&8 \\3&6&7&9\\7 &8&8\\ 9& 9&9
 \end{ytableau}\,\,\,
\hspace{0.3cm}
 \begin{ytableau}
 1&1 & 1&4\\2 &2&2&7 \\3&6&7&9\\8 &8&8\\ 9& 9&9
 \end{ytableau}
\hspace{0.3cm}
 \begin{ytableau}
 1&1 & 1&4\\2 &2&2&6\\3&7&7&9\\8 &8&8\\ 9& 9&9
 \end{ytableau}
\end{align}
The defining properties of $\tilde{S}_2^{\text{hws}}$ imply that $S_1\otimes \tilde{S}_2^{\text{hws}}$ should nullify the non negative operator $\text{Op}^{S_1\otimes \tilde{S}_2^{\text{hws}}}_{\bar{\gamma}_L}=\sum_{q=0}^{k_2-2}E_{L-q-1,L-q}E_{L-q,L-q-1}$, where $k_2$ is the number of rows of $\bar{\beta}_{q_2}$, i.e  $k_2=\text{Max} \{j \mid \bar{\beta}_{q_2}(j) \neq 0 \}$.
Note that if there is just one row in $\bar{\beta}_{q_2}$, then $k_2=1$, the set of SSYT for the expansion of $S_1\otimes \tilde{S}_2^{\text{hws}}$ is reduced to only one SSYT, there is no term in $\text{Op}^{S_1\otimes \tilde{S}_2^{\text{hws}}}_{\bar{\gamma}_L}$ and no need to build it.
For our example, $\text{Op}^{S_1\otimes \tilde{S}_2^{\text{hws}}}_{\bar{\gamma}_L}$ is represented, on the basis of Eq. \ref{basis}, by the following matrix:
\begin{equation} \label{matrix_operator}
\begin{pmatrix} 
2& \sqrt{2} &0 &0 \\ \sqrt{2}  & \frac{17}{5} & \frac{6}{5} & 0  \\ 0 & \frac{6}{5} & \frac{49}{15} & \frac{2\sqrt{2}}{3} \\ 0 & 0 &  \frac{2\sqrt{2}}{3} & \frac{1}{3}
\end{pmatrix},
\end{equation} 
that we have calculated from the matrix elements of the generators $E_{L-q-1,L-q}$ using the GT rules \cite{Gelfand_1950} (cf Appendix \ref{GT_rules_appendix}).
By Gaussian elimination or by diagonalization, one obtains for  $S_1\otimes \tilde{S}_2^{\text{hws}}$ :
\begin{align} \label{subd_coeff1}
\ytableaushort{1114,222,3} \otimes \ytableaushort{9999,888,77,6} &=\frac{-1}{5 \sqrt{3}} \raisebox{1.4ex}{\begin{ytableau}
 1&1 & 1&4\\2 &2&2&8 \\3&7&7&9\\6 &8&8\\ 9& 9&9
 \end{ytableau}}
 +\frac{\sqrt{2}}{5 \sqrt{3}}
 \raisebox{1.4ex}{ \begin{ytableau}
 1&1 & 1&4\\2 &2&2&8 \\3&6&7&9\\7 &8&8\\ 9& 9&9
 \end{ytableau}} \nonumber \\
&-\frac{2\sqrt{2}}{5 \sqrt{3}}
  \raisebox{1.4ex}{\begin{ytableau}
 1&1 & 1&4\\2 &2&2&7 \\3&6&7&9\\8 &8&8\\ 9& 9&9
 \end{ytableau}}
+\frac{8}{5 \sqrt{3}}
 \raisebox{1.4ex}{ \begin{ytableau}
 1&1 & 1&4\\2 &2&2&6\\3&7&7&9\\8 &8&8\\ 9& 9&9
 \end{ytableau}}
\end{align}
Note that in general, the dimension of the nullspace of $\text{Op}^{S_1\otimes \tilde{S}_2^{\text{hws}}}_{\bar{\gamma}_L}$ is equal to the outer multiplicity of $\bar{\gamma}_L$ in $\bar{\beta}_{q_1} \otimes \bar{\beta}_{q_2}$, i.e $T^{\bar{\gamma}_L}_{N_s+1}(q_1,q_2)$.
The coefficients in the right hand side (rhs) of Eq. \ref{subd_coeff1}
are some $U(r_1+r_2+2)\supset U(r_1+1) \otimes U(r_2+1)$ subduction coefficients.

 \underline{Step 3}

 We apply a linear superposition of product of hopping between consecutive sites $E_{q+1,q}$,  to put the $n_2$ cross at the rows specified by the entries of the vector $l_2$.
 We detail here how to obtain such a linear superposition on our example  $(\bar{\beta}_{q_2},l_2)=([4,3,2,1],[3 2])$, in which there are $n_2=2$ cross.
 We proceed cross by cross from the one at the highest row, in order to transform 
  \begin{align}
 S_1 \otimes \tilde{S}_2^{\text{hws}}= \ytableaushort{1114,222,3} \otimes \ytableaushort{9999,888,77,6}
 \end{align}
 into
   \begin{align}
 S_1 \otimes \tilde{S}_2=  \ytableaushort{1114,222,3} \otimes \ytableaushort{9999,885,75,6}.
 \end{align}

\begin{figure*} 
\centerline{\includegraphics[width=1\linewidth]{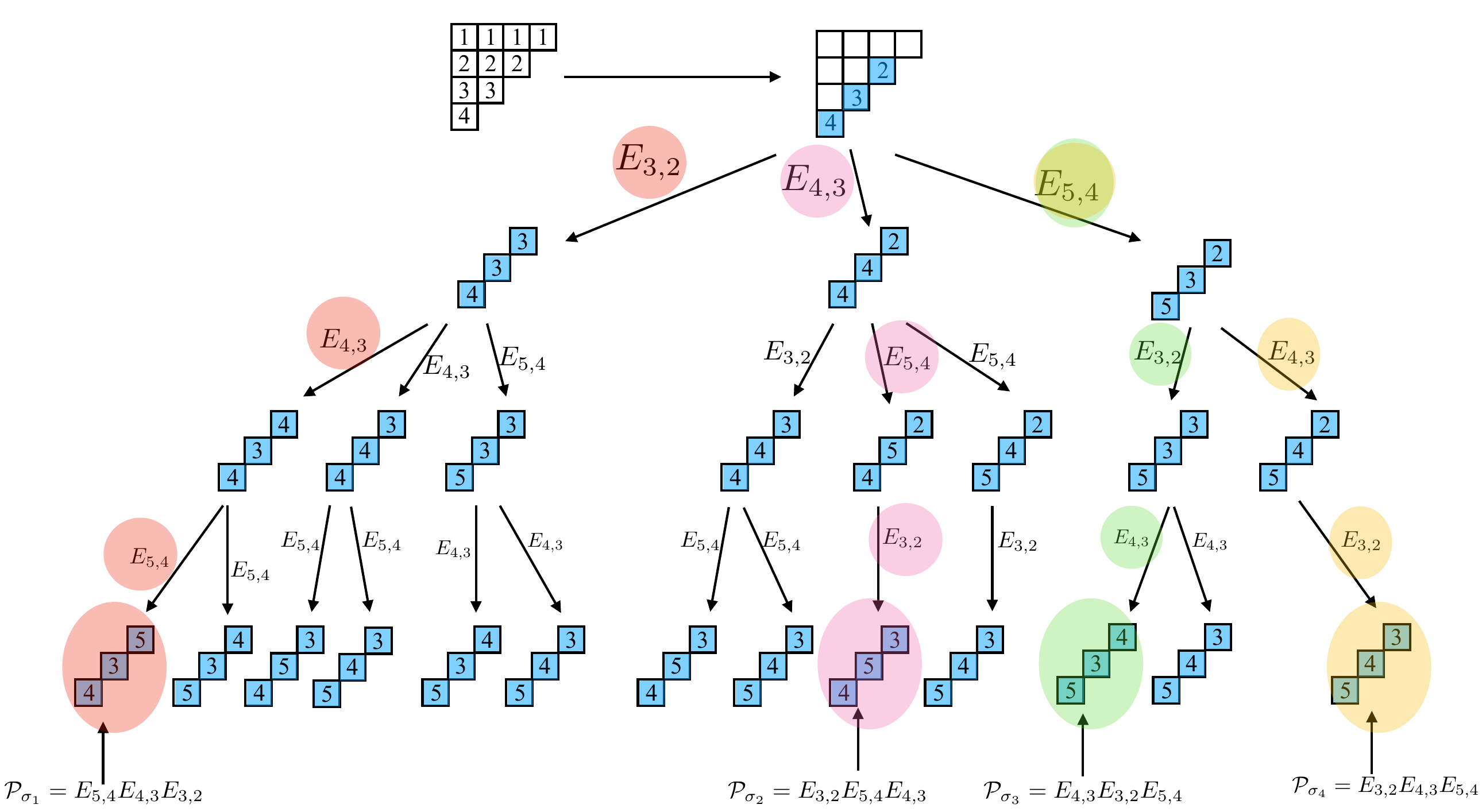}}
\caption{\label{tree} Example of tree generated by the application of operators of the form  $\mathcal{P}_{\sigma} =\prod^{r_2}_{k=l_2(\text{end})} E_{\sigma(k)+1,\sigma(k)}$ for $\bar{\beta}_{q_2}=[4 3 2 1]$, $l_2(\text{end})=2$ and $r_2=4$, where $\sigma$ is a permutation of $\{l_2(\text{end}),l_2(\text{end})+1, \cdots , r_2\}(=\{2,3,4\})$ (cf text for details). 
In particular, each time an operator $E_{\sigma(k)+1,\sigma(k)}$ acts on a SSYT, either it vanishes, either it creates some SSYTs where one number in the {\bf edge boxes} has changed. On a given SSYT, for each row between row $l_2(\text{end})=2$ and row $r_2=4$ inclusive, the edge boxes are the  rightmost  boxes and appear in blue in the figure.
From this tree, our purpose is to select the minimal set of permutations $\{\sigma\}$ to generate the largest variety of SSYT (containing $S_2^{1/n_2}$, cf Eq. \ref{S_2_1_n_2}) appearing on the last line of the tree. To make such a selection, one can scan through the final states (for instance from right to left here) and each time there is a new final state, one selects the corresponding permutation $\sigma$.
}
\end{figure*}

 Let's detail the sequence of computations:
  First, one should determine the coefficients $ \eta (\sigma)$ in the operation 
 \begin{align}
 \mathcal{T}_{S_2^{\text{hws}}}^{S_2^{1/n_2}}&=\sum_{\sigma} \eta (\sigma) \mathcal{P}_{\sigma} \label{trans_T_S2}
 =\sum_{\sigma} \eta (\sigma) \prod^{r_2}_{k=l_2(\text{end})} E_{\sigma(k)+1,\sigma(k)}, 
 \end{align}
  that will be such that  $ \mathcal{T}_{S_2^{\text{hws}}}^{S_2^{1/n_2}} S_2^{\text{hws}}=S_2^{1/n_2}$, 
 which, for our example, reads:
 \begin{align}
 \label{S_2_1_n_2}
 S_2^{1/n_2}=
 \raisebox{1.4ex}{  \ytableaushort{1111,225,33,4}},
  \end{align}
  since the highest row for the first cross in $(\bar{\beta}_{q_2},l_2)=([4,3,2,1],[3 2])$ is row number $l_2(n_2)=l_2(\text{end})=2$.
To determine the coefficients $\eta (\sigma)$ in Eq. \ref{trans_T_S2}, one could consider all the permutations $\sigma$ of $\{l_2(\text{end}),l_2(\text{end})+1, \cdots , r_2\}(=\{2,3,4\})$,
but for tall SSYT $S_2^{\text{hws}}$, with a cross located in the first rows, it won't be efficient, as we will handle a lot of permutations.
Such a number of permutation is $(r_2-l_2(\text{end})+1)!$, which is equal to $3!=6$ in our example, but we will show that one can write $\mathcal{T}_{S_2^{\text{hws}}}^{S_2^{1/n_2}}$ as a sum of only four different terms.
In fact, applied on the highest weight SSYT $S_2^{\text{hws}}$, each hopping term $ E_{\sigma(k)+1,\sigma(k)}$ appearing in a product $\mathcal{P}_{\sigma} =\prod^{r_2}_{k=l_2(\text{end})} E_{\sigma(k)+1,\sigma(k)}$  will just change the numbers in the 
{\bf edges} boxes (cf Fig. \ref{tree}) between row $l_2(\text{end})$ and row $r_2$, in such a way that the selection of the minimal set of permutations $\sigma$ just requires the  generation of the tree shown in Fig \ref{tree}.
At each stage of the tree, we just use the defining constraints of the SSYT for the edge boxes to know whether the product by a hopping term $ E_{\sigma(k)+1,\sigma(k)}$ is possible whether it gives $0$, and which new SSYT it can create.
We then select the minimal number of permutations $\sigma$ such that the products $\mathcal{P}_{\sigma} $ can generate the largest number of different SSYT, which in particular, 
should contain $S_2^{1/n_2}=S_2^{1/2}$.
Such a selection can be made by scanning through the final states and each time there is a new finale state, one adds the corresponding permutation $\sigma$ (cf Fig. \ref{tree}).
Thus for our example, the four products of hopping terms are:
\begin{align}
&\mathcal{P}_{\sigma_1}= E_{5,4}E_{4,3}E_{3,2}   \hspace{1cm} \mathcal{P}_{\sigma_2}=E_{3,2}E_{5,4}E_{4,3}   \nonumber \\
&\mathcal{P}_{\sigma_3}=E_{4,3}E_{3,2}E_{5,4} \hspace{1cm} \mathcal{P}_{\sigma_4}=E_{3,2}E_{4,3}E_{5,4}. 
\end{align}
Finally, we apply the GT rules (cf Appendix \ref{GT_rules_appendix}) to calculate all the $\mathcal{P}_{\sigma} S_2^{\text{hws}}$, and we obtain a linear system to get the coefficients $ \eta (\sigma) $ such that 
Eq. \ref{trans_T_S2} is satisfied. On our example, it gives:
\begin{align}
&\mathcal{P}_{\sigma_1} S_2^{\text{hws}}= \sqrt{\frac{15}{8}}\vert A \rangle + \sqrt{\frac{3}{4}}\vert B \rangle +\sqrt{\frac{9}{8}} \vert C \rangle + \sqrt{\frac{1}{2}}\vert D \rangle, \nonumber \\
&\mathcal{P}_{\sigma_2} S_2^{\text{hws}}=  \sqrt{3}\vert B \rangle +\vert D \rangle, \nonumber \\
&\mathcal{P}_{\sigma_3} S_2^{\text{hws}}= \sqrt{2} \vert C \rangle +\vert D \rangle, \nonumber \\
&\mathcal{P}_{\sigma_4} S_2^{\text{hws}}= 2\vert D \rangle, \hspace{1.5cm} \text{where}\nonumber \\
&\vert A \rangle=S_2^{1/n_2}=  \raisebox{1.4ex}{\ytableaushort{1111,225,33,4}}, \hspace{.5cm} \vert B \rangle=\raisebox{1.4ex}{\ytableaushort{1111,223,35,4}}, \nonumber \\ 
&\vert C \rangle=\raisebox{1.4ex}{\ytableaushort{1111,224,33,5}},\hspace{1.75cm}\vert D \rangle=\raisebox{1.4ex}{\ytableaushort{1111,223,34,5}}, \text{so that} \\
&\mathcal{T}_{S_2^{\text{hws}}}^{S_2^{1/n_2}}=\Big{\{}\sqrt{\frac{8}{15}}\mathcal{P}_{\sigma_1} - \sqrt{\frac{2}{15}}\mathcal{P}_{\sigma_2}-  \sqrt{\frac{3}{10}}\mathcal{P}_{\sigma_3}+\sqrt{\frac{3}{40}}\mathcal{P}_{\sigma_4}\Big{\}}. \nonumber
\end{align}
  
 In order to put the second cross at row number $l_2(n_2-1)(=l_2(1)=3$ here), we first select the minimal number of relevant permutations within 
 $\{l_2(1),l_2(1)+1, \cdots , r_2\}(=\{3,4\})$, before solving the linear system to generate $S_2^{2/n_2}$ from $S_2^{1/n_2}$, as:
  \begin{align}
 S_2^{2/n_2}=
 \raisebox{1.4ex}{  \ytableaushort{1111,225,35,4}}&=  \mathcal{T}_{S_2^{1/n_2}}^{S_2^{2/n_2}}  S_2^{1/n_2} \\
 &=  \Big{\{} \frac{2}{3} E_{5,4}E_{4,3} - \frac{1}{3}E_{4,3}E_{5,4}\Big{\}}  \raisebox{1.4ex}{  \ytableaushort{1111,225,33,4}}. \nonumber 
  \end{align}
  
Cross by cross, we proceed the same way to determine all the coefficients in the operators $ \mathcal{T}_{S_2^{k/n_2}}^{S_2^{k+1/n_2}}$, 
  for $k=0, 1 , \cdots n_2-1$, with $S_2^{0/n_2} \equiv S_2^{\text{hws}}$, to build the product $R_{S_2^{\text{hws}}}^{S_2^{n_2/n_2}}=\mathcal{T}_{S_2^{n_2-1/n_2}}^{S_2^{n_2/n_2}} \cdots  \mathcal{T}_{S_2^{1/n_2}}^{S_2^{2/n_2}}  \mathcal{T}_{S_2^{\text{hws}}}^{S_2^{1/n_2}}$. Then, we replace the numbers $1,2, \cdots r_2+1$ by $L, L-1, \cdots r_1+2$  in $R_{S_2^{\text{hws}}}^{S_2^{n_2/n_2}}$ to obtain
$\tilde{R}_{S_2^{\text{hws}}}^{S_2^{n_2/n_2}}$  and we apply $\tilde{R}_{S_2^{\text{hws}}}^{S_2^{n_2/n_2}}$ on  $S_1 \otimes  \tilde{S}_2^{\text{hws}}  $ to obtain $S_1 \otimes \tilde{S}_2$:
  \begin{align}
S_1 \otimes \tilde{S}_2 = \tilde{R}_{S_2^{\text{hws}}}^{S_2^{n_2/n_2}} S_1 \otimes  \tilde{S}_2^{\text{hws}},
  \end{align}
  where we use the expansion obtained at the end of step $2$ for $S_1 \otimes  \tilde{S}_2^{\text{hws}}  $.

\underline{Step 4 \& 5\& 6}

We perform the steps 1 \& 2 and 3 but with $(\bar{\beta}_{q_3},l_3)$ instead of $(\bar{\beta}_{q_1},l_1)$ and $(\bar{\beta}_{q_4},l_4)$ instead of $(\bar{\beta}_{q_2},l_2)$.

\underline{Step 7}

We finally use (again) the GT rules (cf Appendix \ref{GT_rules_appendix}) on the irrep $\bar{\gamma}_L$ seen as an irrep of the unitary group $U(L=r_1+r_2+2)$ to calculate:
\begin{align}
&\langle \bar{\beta}_{q_3},l_3  \vert \otimes\langle \bar{\beta}_{q_4},l_4 \vert  E^{\bar{\gamma}_L}_{N_s+1,N_s+2} \vert \bar{\beta}_{q_1},l_1\rangle \otimes \vert \bar{\beta}_{q_2},l_2\rangle \nonumber \\ &=\langle S_3 \otimes \tilde{S}_4 \vert  E_{r_1+1,r_1+2} \vert S_1 \otimes \tilde{S}_2 \rangle.
\end{align}
which is equal to $\sqrt{3/8}$ in our example.
 Please note that in such an example, $\bar{\gamma}_L$ appears with outer multiplicity equal to one in both $\bar{\beta}_{q_1} \otimes \bar{\beta}_{q_2}$ and in  $\bar{\beta}_{q_3} \otimes \bar{\beta}_{q_4}$,
 i.e $T^{\bar{\gamma}_L}_{N_s+1}(q_1,q_2)=T^{\bar{\gamma}_L}_{N_s+1}(q_3,q_4)=1$,
 which simplified the previous treatment.
 However, the methodology we have developed is not restricted to outer multiplicity $\leq 1$: let's mention the changes implied by outer multiplicity $T^{\bar{\gamma}_L}_{N_s+1}(q_1,q_2)>1$. 
 In general, $T^{\bar{\gamma}_L}_{N_s+1}(q_1,q_2)$ is the dimension of the nullspace of $\text{Op}^{S_1\otimes \tilde{S}_2^{\text{hws}}}_{\bar{\gamma}_L}$, so that there is a freedom (gauge or ambiguity) in the $T^{\bar{\gamma}_L}_{N_s+1}(q_1,q_2)$ vectors of subduction coefficients in the rhs of Eq. \ref{subd_coeff1} defining the basis of such a nullspace.
 Note that gauge issues also exist when $T^{\bar{\gamma}_L}_{N_s+1}(q_1,q_2)=1$. For instance, in Eq. \ref{subd_coeff1}, one could multiply all the coefficients of the RHS by $-1$. Again, the requirement is to stay consistent (by imposing for instance positive sign for the coefficient of the {\it largest} SSYT of the linear superposition).
  One then requires some consistency in the selection of the basis: the way the coefficients are chosen is fixed for a given set of irreps $\bar{\gamma}_L, \bar{\beta}_{q_1}$ and $\bar{\beta}_{q_2}$, and one introduces an additional index $k_i=1,2, \cdots, T^{\bar{\gamma}_L}_{N_s+1}(q_1,q_2)$, to refer to the $k_i^{\text{th}}$ basis state of such a nullspace, which should be added in both the bra and the ket of reduced matrix elements in Eq. \ref{coeff_ex}.  Steps $3$ and steps $7$ are kept identical. 

\subsubsection{Examples of fittings}
\label{fitting_example}

\begin{figure} 
\centerline{\includegraphics[width=1\linewidth]{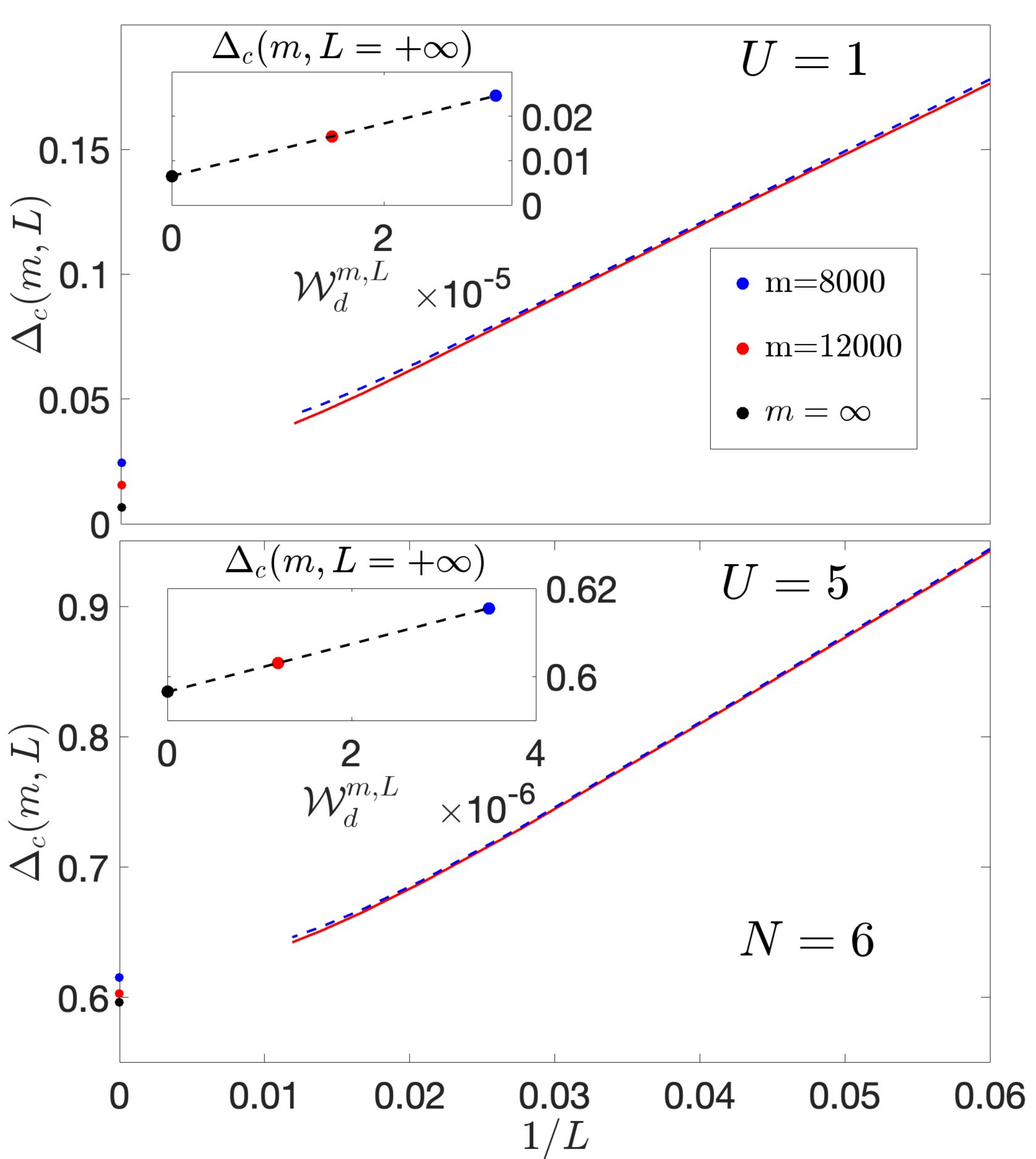}}
\caption{\label{Figure_fitting} Charge gaps $\Delta_c(m,L)$ for N=6 for $m=m_2=12000$ (resp. $m=m_1=8000$) states kept in solid red (resp. dashed blue) lines, for $U=1$ (top) and $U=5$ (bottom) as a function of $1/L$. At fixed $m$, we extrapolate the gaps in the thermodynamical limit $L \rightarrow \infty$, fitting the points $\Delta_c(m,L)$ with a function of the form $a+b/L+d/L^2$ to obtain $\Delta_c(m,L=+\infty)=a$, shown as colored points on the y axis. Insets: values $\Delta_c(m,L=+\infty)$ plotted as a function of the discarded weight
 $\mathcal{W}_d^{m,L}$ for $L=L^{\Delta_c}_{\text{Max}}=84$. We make a linear extrapolation  $\mathcal{W}_d^{m,L}$ to obtain $\Delta_c(m=+\infty,L=+\infty)$ in black, which is tabulated in Tab. \ref{table: table1} and plotted as a function of $U$ in Fig. \ref{gap_charges}.}
\end{figure}

We first show in Fig. \ref{Figure_fitting}, for $N=6$, $U=1$ and $U=5$, the extrapolation of the charge gaps in the $L=+\infty$ limit at fixed $m$, that we denote $\Delta_c(m,L=+\infty)$, and then the extrapolation of the 
resulting limit $\Delta_c(m,L=+\infty)$ in the limit $m\rightarrow+\infty$, which amounts to sending the discarded weight $\mathcal{W}_d^{m,L}$ to 0 (cf inset of Fig. \ref{Figure_fitting}),
in order to obtain $\Delta_c(m=+\infty,L=+\infty)$.

\begin{figure} 
\centerline{\includegraphics[width=1\linewidth]{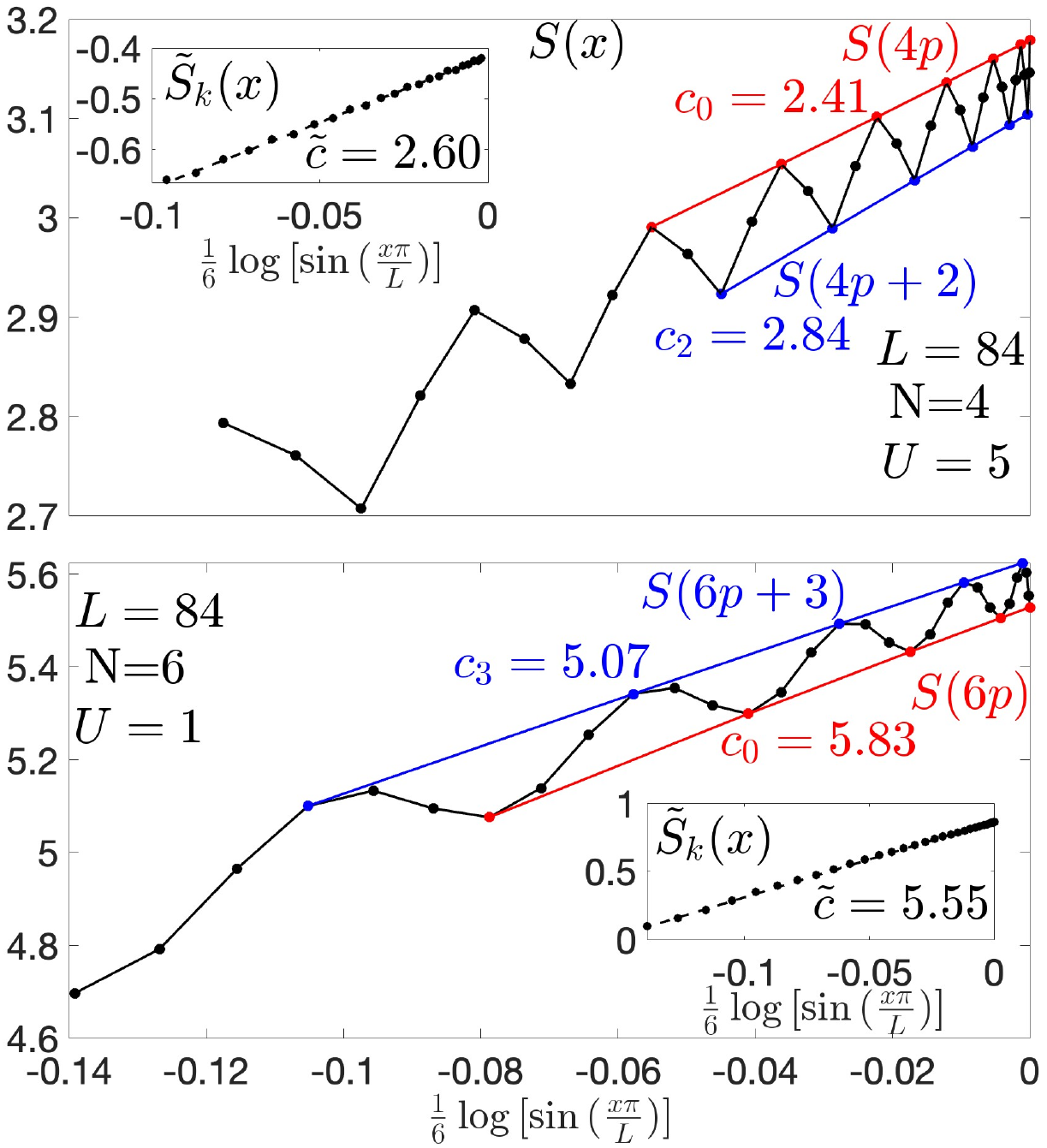}}
\caption{\label{Figure_fitting_entropy} Entanglement entropy $S(x)$ as a function of the logarithm of the conformal distance $\frac{1}{6} \log\Big{[}\sin\Big{(}\frac{\pi x}{L}\Big{)}\Big{]}$ shown for an open chain of $L=84$ sites for N=4, $\,U=5$ and $m=10000$ states kept in the top figure, and for N=6, $\,U=1$ and $m=16000$ in the bottom figure. 
The position $x$ is $x=N\times p+q$ (with $q=0,1,2,\cdots N-1$), giving rise to a most $N$ different straight lines. We show in red $q=0$ and in blue $q=N/2$. Insets: we alternatively consider  $\tilde{S}_k(x)=S(x)+k \langle E_{x,x+1} + \text{h.c}\rangle$, with the best $k$ to remove the Friedel oscillations, Here, $k=2.842$ for N=4 and $k=2.505$ for N=6. The three values $c_{\text{Floor(N/2)}}$ $\tilde{c}$ and $c_{0}$, appear in the current Figure close to the corresponding slopes. They are also tabulated in Tab. \ref{table: table1}, and agree with the field theory predictions (cf main text for details).}
\end{figure}

Secondly, we give in Fig. \ref{Figure_fitting_entropy} the entanglement entropy $S(x)$ as a function of the logarithm of the conformal distance $\frac{1}{6} \log\Big{[}\sin\Big{(}\frac{\pi x}{L}\Big{)}\Big{]}$  for $L=84$ sites and $N=4$ and $N=6$.
Due to Friedel oscillations, there are different slops $c_q$ (with $q=0,1,2,\cdots N-1$) giving rise to a most $N$ different straight lines.
We can also use the bond $\langle E_{x,x+1} + \text{h.c}\rangle$ to remove the Friedel oscillations (cf insets of Fig. \ref{Figure_fitting_entropy}).

 \bibliography{DMRG_FHM.bib}
\end{document}